\documentclass[fleqn,usenatbib]{mnras}

\usepackage{mathtools}						%% BJ: standard \coloneqq or \vcentcolon for :=
\usepackage[T1]{fontenc}
\usepackage{ae,aecompl}

\usepackage{array}
\usepackage{lmodern}
\usepackage{graphicx}
\usepackage{amsmath}	% Advanced maths commands
\usepackage{amssymb}	% Extra maths symbols
\usepackage{bm} % for bold mat symbols
\usepackage{txfonts}    % load this after amsmath to avoid integral sign error
\usepackage[dvipsnames]{xcolor}
\usepackage{multirow} % multiple rows
\usepackage[normalem]{ulem} % for strikethrough of text (remove when finished)
\title[Topological halo bias]{Topological bias: \\ How haloes trace structural patterns in the cosmic web}

\author[R. Bermejo et al.]{Raul Bermejo$^{1}$\thanks{E-mail: raulbv.personal@gmail.com},
Georg Wilding$^{1,2,3}$\thanks{E-mail: wilding@astro.rug.nl},
Rien van de Weygaert$^{1,3}$\thanks{E-mail: weygaert@astro.rug.nl},
Bernard J.T. Jones$^{1}$,
\newauthor
Gert Vegter$^{2,3}$,
Konstantinos Efstathiou$^{4}$
\\
$^{1}$Kapteyn Astronomical Institute, University of Groningen, PO Box 800, 9700 AV Groningen, The Netherlands\\
$^{2}$Bernoulli Institute for Mathematics, Computer Science and Artificial Intelligence, University of Groningen,\\\quad PO Box 800, 9700 AV Groningen, The Netherlands\\
$^{3}$Centre for Data Science and Systems Complexity, University of Groningen, PO Box 800, 9700 AV Groningen, The Netherlands\\
$^{4}$Division of Natural and Applied Sciences and Zu Chongzhi Center for Mathematics and Computational Science,\\\quad Duke Kunshan University, No. 8 Duke Avenue, Kunshan 215316, Jiangsu Province, China
}

\date{Accepted XXX. Received YYY; in original form ZZZ}

\pubyear{2022}

\begin{document}
\label{firstpage}
\pagerange{\pageref{firstpage}--\pageref{lastpage}}
\maketitle

% Abstract of the paper (250 words max), currently 249, (had to be shortened mich more from 335)
\begin{abstract}
We trace the connectivity of the cosmic web as defined by haloes in the Planck-Millennium simulation using a persistence and Betti curve analysis. We normalise clustering up to the second-order correlation function, and use our systematic topological analysis to correlate local information and properties of haloes with their multi-scale geometrical environment of the cosmic web (elongated filamentary bridges and sheetlike walls). We capture the multi-scale topology traced by the halo distribution through filtrations of the corresponding Delaunay tessellation. The resulting nested \emph{alpha shapes} are sensitive to the local density, perfectly outline the local geometry, and contain the complete information on the multi-scale topology. We find a remarkable linear relationship between halo masses and topology: haloes of different mass trace environments with different topological signature. This is \emph{topological bias}, an environmental structure bias independent of the halo clustering bias associated with the two-point correlation function. This mass-dependent linear scaling relation allows us to take clustering into account and determine the overall connectivity from a limited sample of galaxies. The presence of topological bias has major implications for the study of voids and filaments in the observed distribution of galaxies. The (infra)structure and shape of these key cosmic web components will strongly depend on the underlying galaxy sample. Their use as cosmological probes, with their properties influenced by cosmological parameters, will have to account for the subtleties of topological bias. This is of particular relevance with the large upcoming galaxy surveys such as DESI, Euclid, and the Vera Rubin telescope surveys.

\end{abstract}


%%%%%%%%%%%%%%%%%%%%%%%%%%%%%%%%%%%%%%%%%%%%%%%%%%

%%%%%%%%%%%%%%%%% BODY OF PAPER %%%%%%%%%%%%%%%%%%

%########################################################
%                         INTRODUCTION
%########################################################
\section{Introduction}
\label{sec: Introduction}
Considering the Universe at a Megaparsec scale, the distribution of matter becomes anisotropic and inhomogeneous, and a highly geometric and connected structure emerges. This structure is known as the \textit{cosmic web}~\citep{bond1996filaments,weygaert2008clusters,cautun2014}. The cosmic web is, first and foremost, defined through its dark matter distribution. This distribution of densities is easy to obtain if one works on simulations, and allows a straightforward analysis of many of its features, as well as, for example, an identification of its constituent components. Observationally, the cosmic web is studied mostly through the spatial distribution of galaxies. It is, however, not clear in how far the structure traced by these galaxies is representative for the complete structure of the cosmic web.

We address this issue here: in this study, we focus on the cosmic web as traced by dark matter haloes, where we come upon the same problem -- it is unclear how exactly the spatial pattern of dark haloes relates to the underlying matter distribution. As the spatial pattern of the cosmic web is determined by its constituents, which in turn are defined on the basis of their shape and mutual spatial connections, this naturally leads us to the field of topology. To shed light on the multi-scale nature of the topology of the cosmic web traced by the population of dark haloes, we invoke persistent homology to decode the intricate features, patterns, and connectivity of the cosmic web.

Irrespective of the defining components, four notable qualities characterise the cosmic web. These are its anisotropy, the asymmetry between overdense and underdense regions, the connectivity between different morphological features, and its multi-scale nature. The anisotropy of this web-like structure expresses itself through a rich morphology dominated by elongated filaments and tenuous walls. The asymmetry concerns the contrast between the overdense regions -- clusters, filaments and a fair fraction of walls -- and underdense regions. Together with cosmic walls and filaments, we see structure and morphological features over a wide range of scales, a direct manifestation of the bottom-up hierarchical buildup of the cosmic web and its structural components. Detailed theoretical discussions on the hierarchical evolution of the cosmic web have elucidated the intricate multiscale nature of the processes involved \citep[see e.g.][]{bondmyers1996a,hanami2001,sheth2002,shethwey2004,shen2006,weygaert2008clusters,cadiou2020}.

\subsection{The connectivity and topology of the cosmic web}
%%% -------------------------------------------------------------
Of particular importance in the context of the topology of the cosmic matter distribution is the weblike network into which these morphological elements are organised. The resulting distinct connectivity finds its expression in clusters of galaxies at the junction of filaments, filaments at the intersection and boundaries of walls, and filaments and walls defining the boundaries of the large near-empty voids~\citep{Shandarin1989,bond1996filaments,weygaert2008clusters,aragon2010multiscale,aragon2010spine,sousbie2011persistent,sousbie2011persistent2,cautun2014,libeskind2018,codis2018connectivity}. In recent studies, the connectivity of the cosmic web in terms of the number of filaments that connect to a
node has received considerable interest. This aspect of connectivity was first investigated by~\citet{aragon2010spine} on the basis of the MMF formalism~\citep{aragon2007mmf}. The resulting dependence
on the mass of the cluster nodes has recently been confirmed in a more profound and thorough analysis~\citep{codis2018connectivity}.

The connectivity is a key factor in the topological analysis of the present study. To describe and characterize
the multi-scale connectivity of the cosmic web we turn to the mathematical characterisation of connectivity: topology and persistent homology~\citep{munkres1984,edelsbrunner2010computational}. In the study preceding the present one,
\cite{wilding2021}, we established the intimate link between the geometric structure of the cosmic web and its morphological elements and the topological signatures of the cosmic mass distribution. The surprising finding was that not only the structure of the cosmic web finds its expression in the topological parameters - Betti numbers and persistence diagrams - of the mass distribution but that even the key topological transitions such as the establishment of a connecting filamentary network are closely related to the dynamics of structure formation. Most telling is that the filaments start to define a percolating network at the density level corresponding to the gravitational binding of mass concentrations. On the basis of this intimate relationship between structure and topology, in the remainder of the
present study we take this as an established and known fact. 

The present study adresses the connectivity of the cosmic web as sampled by dark haloes,  More specifically, we use Betti numbers~\citep{poincare1892betti} and persistence diagrams to study the connectivity of the the spatial distribution of dark haloes. In the present study, we do so by assessing the spatial distribution of dark matter haloes in a state-of-the-art dark matter simulation marked by a large cosmic volume and vast dynamic range. The Planck Millennium (P-Millennium hereafter, see~\citet{Punya2018} and~\citet{Baugh2019}) contains $7.7\times10^7$ of these haloes.

Here we focus on the topology, and in particular connectivity, of the cosmic mass distribution as traced by the discrete population of haloes. The dark matter haloes, and implicitly the galaxies that reside in their interior, populate the components of the cosmic web. It means that the spatial pattern outlined by dark matter haloes constitutes a diluted and possibly biased reflection of the underlying web-like dark matter distribution. We investigate the dependence of the web-like distribution of dark haloes as a function of their intrinsic properties, in particular that of their mass. This involves an analysis of the patterns and topological features and characteristics -- Betti numbers and persistence -- of the web-like structure delineated by dark matter haloes. We assess these properties for the spatial distribution of haloes in six mass ranges. The aim is to decode the topological embedding of the haloes, i.e., the relation between the morphological components of the cosmic web traced by haloes, their topological characteristics and connectivity, and the mass of the haloes.

P-Millennium provides a very useful halo catalogue for the topological analysis presented in this
study because of its broad mass range of haloes. 
Haloes vary in masses of $\log_{10}\left(M\right) = \left[10.5, 13\right] \ h^{-1}M_{\odot}$, allowing to resolve any topological differences between haloes more precisely. To complete such a task, we divided the halo catalogue into six mass bins. Throughout the rest of the paper, these bins are labelled as HA, HB, HC, HD, HE and HF. Their associated mass ranges and number of haloes are shown in Table~\ref{tab:Classification Table}.

\begin{table}
\centering
\begin{tabular}{@{}llll@{}}
\hline
Halo Population & $\log_{10}$(M) \ {[}$h^{-1}M_{\odot}${]}   & N (number of haloes)     \\ \hline
HA              & {(}10.5,11{]}             & $\simeq 7.05\times10^6$ \\
HB              & (11,11.5{]}               & $\simeq 2.67\times10^6$ \\
HC              & (11.5-12{]}               & $\simeq 9.69\times10^5$ \\
HD              & (12,12.5{]}               & $\simeq 3.47\times10^5$ \\
HE              & (12.5,13{]}               & $\simeq 1.21\times10^5$ \\
HF              & (13,13.5{]}           & $\simeq 4.05\times10^4$ \\ \hline
\end{tabular}
\caption{\textbf{Halo sub-populations in the P-Millennium simulation}. For each halo class we include the logarithmic mass range and the approximate number of haloes for each population.}
\label{tab:Classification Table}
\end{table}

As a basis to analyse the topology of these haloes we use a distance-based filtration. In this kind of filtration, haloes and clusters of haloes form connections if closer than a specific distance. This way of defining connections on a point-based distribution (the dark matter haloes) is formalised through a Delaunay tessellation~\citep{delone1934sphere} and the resulting simplicial complex, which represents the halo structure as a collection of simplices. The simplices are naturally adaptive to the local density and geometry. Choosing a specific filtration length yields a uniquely defined scale-dependent 
subset of the full Delaunay complex, the so-called \emph{alpha shapes}, introduced by Edelsbrunner and collaborators~\citep{Edelsbrunner1983, edelsbrunner1994, edelsbrunner2010computational}. The \emph{alpha complex}, the nested sequence of alpha shapes, perfectly traces the multi-scale connections established by the halo distribution, while considering that the resulting connections occur on all scales. The sequence forms a  direct and unambiguous representation of the multi-scale topological structure of the distance field. The derived persistence diagrams and Betti numbers characterise the underlying homology of the alpha shapes of the halo distribution in this simulation. Through the use of persistence computed from the alpha shape filtration, we establish a natural description of the multi-scale nature of the halo distribution.

The applications of persistent homology~\citep{edelsbrunner2010computational}, Betti numbers~\citep{poincare1892betti}, and topological data analysis in general has seen a large increase in number and popularity in recent years~\citep[for a review see][]{wasserman2018}. We see the popularity also in a diverse range of application fields, ranging from brain research~\citep{petri2014,reimann2017} to materials science~\citep{hiraoka2016}, and in recent years also to cosmology and astrophysics. Over the past decade, an increasing number of cosmological studies have invoked persistent homology in an attempt towards quantifying and classifying the complex weblike patterns in the cosmic matter and galaxy distribution~\citep{weygaert2010alphashapetopology,weygaert2011alpha,sousbie2011persistent,sousbie2011persistent2,shivashankar2016felix,kimura2016,pranav2017topology,kono2020,feldbrugge2019,biagetti2020}. Additional interesting applications concern the investigation of the evolving topology of the reionization network~\citep{elbers2019, Thelie2021, giri2021} and the exploration of the structure and topology of magnetic fields in the interstellar medium~\citep{makarenko2018}.

\subsection{Environmental influences on the formation and evolution of haloes and galaxies}
%%% -------------------------------------------------------------
The dependence of the properties of haloes of galaxies, and of their formation and evolution, on the large-scale environment is a vigorous issue of study and discussion in the cosmological community. The first direct indication for a major systematic influence is the discovery by~\citet{dressler1980} of the density-morphology relation. Early-type galaxies are predominantly found in high density regions, in particular clusters of galaxies, while late-type galaxies form the majority of galaxies in more moderate density environments. A particularly nice illustration of this in the context of the web-like galaxy distribution is the segregation between ellipticals and spirals in the Pisces-Perseus supercluster~\citep{giovanelli1986}.

There are indications for a significant impact of the large-scale environment on a range of properties of galaxies and haloes. Particularly outstanding is the relation between galaxy properties and that of density of the environment, of which the morphology-density relation is the most well-known representative \citep{dressler1980}. The influence of the location in the various morphological environments in the cosmic web -- clusters, filaments, walls and voids -- on halo and galaxy properties appears to be more subtle. Also it remains an as yet unsettled issue whether the local density is the sole dominant factor, or whether the morphological nature of the environment also represents a significant supplementary or modulating influence~\citep{Goh2019,hellwing2021}. However, to some extent this concerns an ill-defined dichotomy: the density of the environment is augmented by the anisotropy of its gravitational contraction and collapse~\citep{icke1973,bondmyers1996a,bertjain1994}. 

Many studies indicate that the paramount factor determining the physical nature of a galaxy and its dark matter halo is the mass of the halo.  The mass of haloes is itself intimately linked to their environment in the cosmic web: the halo and galaxy mass functions are strongly determined by whether the haloes reside in filaments, walls, voids or cluster nodes~\citep{cautun2014, Punya2018, punya2019}. While the number density of haloes in filaments, harbouring $~ 50\%$ of dark matter, haloes and galaxies, is somewhat higher than the average density, it is much lower in walls and voids. The mass function in walls and voids is also shifted considerably to lower masses: walls, and even more so voids, are populated by much smaller haloes and galaxies than those populating the filaments, while cluster nodes contain the most massive haloes and galaxies. This is also reflected in the dependence of halo clustering as well as halo bias on halo mass~\citep{Yang2017}, and/or their proximity to morphological features of the cosmic web~\citep{Kraljic2018}\footnote{The definition of proximity to a filament in the cosmic web is to some extent predicated by the particular classification technique used. A method that allows the identification of filaments of varying thickness would relate galaxy properties to the thickness and prominence of the filament.}.

Arguably, the most outstanding influence of the cosmic web on the formation and evolution of galaxies is that of rotation. The tidal force field induced by the evolving inhomogeneous mass distribution is responsible for the anisotropic collapse of filaments and walls, as well as for the torquing of contracting mass clumps~\citep{hoyle1949,peebles1969,white1984,porciani2002b,schafer2009}. This induces a significant correlation between the direction of filamentary ridges and the spinning direction of collapsed haloes~\citep{leepen2001,porciani2002b}. The alignment between filaments and galaxy spin has also been found in observations~\citep{jones2010,tempel2013,vandeSande2021}. Starting with the work by \citet{aragon2007thesis}, a large number of studies have shown that the halo spin alignment with respect to large-scale filaments is mass-dependent~\citep{aragon2007thesis,hahn2007b,zhang2009,codis2012,trowland2013,wang2017,codis2018,Punya2018}, which may be an indication for the influence of anisotropic inflow along the filamentary arteries ~\citep{Punya2018,punya2019,punya2021,lopez2021}. The reality of this effect has recently been confirmed by observations~\citep{tempel2013,welker2020}. 

A major additional influence of the cosmic web on the properties of haloes and galaxies is that on their growth and assembly history. A range of studies have indicated that the growth of haloes is significantly influenced by the cosmic web environment. The relation between halo growth and cosmic web environment is partially a reflection of the tidally directed dynamical evolution of and anisotropic inflow on to the haloes~\citep{aragon2007,dalal2008,hahn2009,porciani2017,musso2018,Paranjape2018,zhang2021}. The result is a halo assembly history and time that is modulated by the cosmic web, and translates into a halo and galaxy bias known as assembly bias~\citep{Gao2005,Wechsler2006,dalal2008,Mao2018}. \citet{salcedo2020} found observational evidence for such bias in the SDSS survey, while it may also be reflected in the relation between the brightness of galaxies and their level of clustering as has been established for the GAMA survey~\citep{Jarrett2017} and SDSS survey~\citep{Paranjape2018}.

\subsection{Topological Bias: Identification \& quantification}
%%% -------------------------------------------------------------
%%% BJ:
%%% I had great difficulty expressing this in a simple way.  
%%% Please check that it makes sense and correct any ambiguities if you can.
The central question of the present study is whether the topological patterns and connections we detect through persistent homology \emph{also} involve a dependence on the intrinsic properties of the tracing haloes. If so, this implies a dependence that we define as \emph{topological bias}.  

Consider the subsample defined by the more massive haloes: it is relatively sparse and has lower density than samples of lower mass haloes.   This situation is directly reflected in the larger scale length of the two-point correlation function for the different mass class haloes.  The different mass samples have greatly different density. The detection of a mass-dependent topological bias requires that we compensate for this.  To achieve this compensation, the spatial scales of the topological features in a given mass class are rescaled using the correlation length of that class. After this renormalisation the connected structures of halo populations of different mass ranges (and accordingly vastly different number densities) can be compared and investigated on scales of similar length and \emph{equal clustering}. 

Accordingly, before running the topological analysis we renormalise the scale of each sample relative to its clustering lengthscale $r_0$. The details of this analysis are presented in section \ref{subsec: Re-scaling}.

Deviations in topological properties of the renormalised halo distribution are implicit indications and manifestations of the presence of higher order clustering. They would elude detection through a two-point correlation function analysis, and are a direct indication for complex topological influences. 

We observe that haloes in different mass ranges trace specific structural patterns, characterised by topological features of varying dimension, at different \emph{normalised} length scales. This is a direct indication and compelling evidence for the presence of a bias that involves nontrivial contributions by higher order clustering terms.  This topological bias is found to have a simple dependence on the mass of the dark matter haloes that define the structure.  Moreover, and perhaps surprisingly, we find a clear systematic behaviour of the topological bias: there is a strict relations between the normalised topological parameters and the characteristics of the halo population. 

%%% BJ: The above paragraph is intended to replace this piece of text
%%% Feel free to reinstate original (with a ffew minor edits perhaps).
% The detection of a mass-dependent topological bias requires that we compensate for the effects of second-order clustering. The sparser samples of massive haloes naturally involve a lower spatial resolution and a dilution of the spatial pattern they trace. Tenuous features like filamentary tendrils and subvoids may not be traceable anymore. It will translate into a significant effect on the topological features that can be resolved. To a large extent these effects are reflected in the two-point correlation function. A major point of interest is whether there there are significant additional topological signatures. 

\subsection{Implications of topological bias}
%%% -------------------------------------------------------------
Topological bias addresses the question in how far the morphology of voids and filaments, and the corresponding tunnels, depend on the nature of the halo populations. It means that the cosmic web traced by haloes in different mass ranges exhibits differences in the corresponding topological characteristics, such as the connectivity properties. More generally, it also involves aspects such as volume, shape and substructure of these features.  In other words, it implies a systematic dependence of the spatial structure of the cosmic web on the specifics of the halo population that is used as its tracer. Moreover, when normalised and compensated for two-point clustering, topological analysis reveals that the structure of the cosmic web  traced by the different halo populations displays fundamental differences. In other words, filaments and voids for different halo populations do have significant differences. These go further than simple self-similarity and includes effects of substructure, connectivity and shape.

The innately multi-scale nature of persistent topology provides new insights into the hierarchical nature of the structure formation process. Structural features are the product of a process marked by a rich assembly history. This tends to leave traces in the (sub)structure of such features, to which the multi-scale topological analysis in terms of homology and persistence is very sensitive. We therefore expect such an analysis to highlight the link between the assembly history and cosmic environment. This may imply an intricate and close relationship between the topological bias which we find and assembly bias~\citep{Gao2005, Wechsler2006, Mao2018}.

Observing topological bias will have a substantial impact on the information on the cosmic web that can be gained from cosmological studies. The detection and incorporation of topological bias in the analysis of cosmological data sets is crucial if one seeks to exploit higher levels of structural complexity in the distribution of galaxies and galaxy haloes. This concerns important aspects such as the properties of the void population, that of the filamentary spine of the cosmic web and its branching network of tendrils. The visually more directly accessible description in terms of topological factors, such as Betti numbers and  persistence, also provides a considerably better appreciation of the nature of the contributing higher order clustering terms. Topological bias may also imply complications when seeking to use the observed galaxy distribution to extract information on the underlying dark matter distribution and on general cosmological parameters. The presence of topological bias reveals systematic differences between the structural pattern outlined by the dark matter distribution and that by galaxies. 

\subsection{Programme and outline}
%%% -------------------------------------------------------------
To unravel the relationship between the topology of the galaxy distribution and the underlying dark matter distribution, we first have to establish a systematic, preferably quantifiable, relationship between the topological characteristics of the spatial patterns outlined by the various populations. The present study limits itself to an investigation of topological properties of populations of dark matter haloes in the P-Millennium simulation, in a subsequent study we will extend this to that of the simulated galaxy distribution in the IllustrisTNG simulations. In a final stage, we will extend our analysis to the observed galaxy distribution. A range of available redshift surveys, e.g. the 2dFGRS, SDSS, GAMA, 2MASS, and VIPERS~\citep{colless2003,tegmark2004cosmological,gama2009,huchra20122mass,vipers2013}, already produced detailed three-dimensional maps that revealed the presence of a rich web-like morphology and topology. The very recent Year 3 data release of DES~\citep{des2021} and the upcoming EUCLID survey~\citep{euclid2011} as well as the DESI experiment~\citep{desi2016,desi2019} will provide ample opportunity for the identification of topological features to analyse their multi-scale (sub)structure, and for applications of the toolset of topology and persistent homology described in the present study.

In this paper we extend the earlier work of the group laid out in~\citet{weygaert2011alpha},~\citet{nevenzeel2013triangulating},~\citet{pranav2017topology},
~\citet{pranav2019unexpected},~\citet{pranav2019topology},~\citet{feldbrugge2019} and~\citet{wilding2021}. It represents a key step in enabling the formalism developed in these previous studies towards one suited for the analysis of galaxy redshift surveys. To this end, this paper is structured as follows: in Section~\ref{sec: Theory}, we introduce the formalism of persistent homology, which is necessary to understand the construction of alpha shapes that yield the persistence diagrams and Betti curves. In Section~\ref{sec:halopopulation} we describe how we obtained the halo catalogue that was used in the study. In this section, we also motivate the mass binning we applied to the catalogue and describe some of its halo properties, such as spatial distribution and clustering. Section~\ref{sec: Results} presents the results of this study, mainly the intensity persistence diagrams and Betti curves which aim to support our claim of a topological bias.
Moreover, in this section we discuss the implications of our intensity persistence diagrams and Betti curves, and explain how these results provide compelling evidence for the existence of a topological bias. Finally, in Section~\ref{sec: Conclusion} we summarise the results of this paper and indicate some future research prospects following this work.

%########################################################
%               TOPOLOGY AND ALPHA SHAPES      
%########################################################
\section{Topology, persistence and alpha shapes} \label{sec: Theory}
%%% BJ: moved to a later place to coincide roughly with first citation
%\begin{figure*}
%\centering
%\includegraphics[width=0.9\textwidth]{figures/1_Evolution_AlphaShape_HE_new.pdf}
%\caption{
%\BJ{\textbf{Alpha shape of a slice of haloes in the P-Millennium.} This plot represents the evolution of a two-dimensional alpha shape for a sample of haloes with a mass range of $\log_{10}\left(M\right) = \left(12.5,13\right] h^{-1}M_{\odot}$ in a slice of $30 h^{-1}$Mpc from the halo catalogue. Different panels are associated with different values of $\alpha$, representing different stages in the evolution of the alpha shape. From left to right in each row: $\alpha = 0 \ h^{-1}$Mpc, $\alpha = 3 \ h^{-1}$Mpc, $\alpha = 6 \ h^{-1}$Mpc, $\alpha = 8 \ h^{-1}$Mpc, $\alpha = 10 \ h^{-1}$Mpc, $\alpha = 14 \ h^{-1}$Mpc. 
%These plots show how, as we increase the value $\alpha$, topological features (such as tunnels) are born and disappear and the haloes become more connected. The last panel represents an almost complete version of the Delaunay tessellation.}
%}
%\label{fig:1_AlphaShape_halos}
%\end{figure*}
%

Topology is the field of mathematics that studies properties which are preserved under continuous deformations (such as bending and stretching, but not gluing). Connectivity is a key \emph{topological invariant}~\citep{Sutherland} which studies mathematical features in a given point distribution. Formally, these topological features are structures that prevent the space to be continuously deformed into a single point. The canonical probe of connectivity is that of counting topological features, and establish their relationship with neighbouring features. 

The most outstanding virtue of our analysis is that it  allows us to evaluate the prominence of the \emph{global} topology of the different topological features as a function of spatial scale, density or other relevant factors. This is of major importance for our understanding of the structure of the cosmic web, and of its hierarchical assembly. 

We are also interested in a direct assessment of the dependence of the structure of the mass distribution on spatial scale as inferred from the observed sample of objects. The multi-scale nature of the matter distribution would follow from the assessment of changes in the field structure as a function of distance (level).
  
\subsection{Topological features}
%%% -------------------------------------------------------------
In the context of this study, we distinguish three different topological features. These are associated with (super)clusters, filaments and voids. Each of these features correspond to a well-defined topological class with corresponding dimensionality. Throughout the rest of the paper, we will be referring to (super)clusters, filaments -- and their associated tunnels -- and voids as \emph{topological features}.

Formally, the dimensionality is associated with the type of topological hole that each feature corresponds to. In the context of the present study, zero-dimensional holes correspond to the space separating two disconnected (super)clusters, a one-dimensional hole is associated with a closed loop of filaments, and a two-dimensional hole corresponds to a void bounded by a closed surface. It is the study of these topological holes and their boundary that provides a formal  notion of connectivity in the distance field that we study. For example, zero-dimensional holes can also refer to the space separating two \emph{(super)clusters} of haloes.

\subsection{Homology}
%%% -------------------------------------------------------------
The formalism of homology allows us to probe the number of zero-, one- and two-dimensional structural features and their connectivity in the underlying matter distribution. The number of the various topological features is expressed in terms of \emph{Betti numbers}~\citep{poincare1892betti}. In proper mathematical formulation, the $d$-th Betti number $\beta_d$ corresponds to the rank of the $d$-th homology group. An in-depth discussion of the theory of homology groups lies outside the scope of this paper. For the interested reader, we refer to~\citet{edelsbrunner2010computational,zomorodian2005,vegter2004,rote2006computational} and \citet{robins2006,robins2015} for a general introduction of homology theory and its application in computational topology. For the more specific case of Betti numbers and persistent homology in the context of the density field of the cosmic web, we refer to, for example, ~\citet{weygaert2010alphashapetopology},~\citet{weygaert2011alpha},~\citet{sousbie2011persistent},~\citet{pranav2017topology},~\citet{pranav2019topology},~\citet{xu2019finding} and~\citet{wilding2021}.

\subsection{Betti numbers and cosmic web topology} \label{subsec:bettinumbers}
%%% -------------------------------------------------------------
Intuitively, the Betti numbers count the number of topological features in the space. That is, in the context of haloes outlining the cosmic web, $\beta_0$ counts the number of disconnected clusters of haloes, $\beta_1$ counts the number of filamentary loops enclosing independent tunnels and $\beta_2$ counts the number of shells enclosing isolated voids. More precisely, the $d$-th Betti number corresponds to the number of $d$-dimensional holes of the sampled space.  

Many studies have probed the topology of the cosmic mass and galaxy distribution with the \emph{Euler characteristic} $\chi$~\citep{Euler1758,adler1981,gott1986sponge, Hamilton1986,park2013,pranav2019topology}, often via the directly related \emph{genus}. The Euler characteristic represents a profound topological invariant, which finds itself at the junction of several branches of mathematics, including homology and simplicial topology~\citep[see][]{adler1981,adler2010,pranav2019topology}. The Euler characteristic of a three-dimensional
set is the number of its connected components, minus the number of its tunnels, plus the number of voids it contains. It is also a geometric quantity via the Gauss-Bonnet theorem, the deep and seminal
expression going back to Euler~\citep{Euler1758}. Requiring both Differential and Algebraic Topology, it shows that the Euler characteristic also has a geometric interpretation, and is associated with the integrated Gaussian curvature of a manifold. In fact, together with other quantities related to volume, area and length, the Euler characteristic forms a part of a more extensive geometrical descriptions in terms of \emph{Minkowski functionals} or {Lipschitz-Killing curvatures}. These have been invoked in a cosmological context in a range of studies~\citep[see e.g][]{mecke1994,schmalzing1997,schmalzing1998,sahni1998,schmalzing1999disentangling,Kerscher2000}. 

There is a profound relationship between the homology characterisation in terms of Betti numbers and the Euler characteristic~\citep[see][]{pranav2019topology}. The \emph{Euler-Poincar\'e} formula states that the Euler characteristic is the alternating sum of the Betti numbers~\citep[also see e.g.][]{weygaert2011alpha,pranav2017topology,pranav2019topology},
\begin{equation}
\chi = \beta_0 - \beta_1 + \beta_2 \,.
\end{equation}
\noindent This expression immediately reveals that the Betti number characterisation of the web-like halo distributions represents a more elaborate and visually insightful description of the topology of the cosmic web than that quantified only in terms of the Euler characteristic or genus. Visually imagining the 3D situation as the projection of three Betti numbers on to a one-dimensional line, we may directly appreciate that two manifolds that are branded as topologically equivalent in terms of their Euler characteristic may actually turn out to possess intrinsically different topologies when described in the richer language of homology. Evidently, in a cosmological content this will lead to a significant increase of the ability of topological analyses to discriminate between different cosmic structure formation scenarios. 

\subsection{Multi-scale topology \& filtrations} \label{subsec:multiscale}
%%% -------------------------------------------------------------
While the Betti number analysis provides a global inventory of structural and topological changes as a function of scale, homology entails a considerably richer source of information with respect to the multi-scale character of the mass distribution. It allows the identification and characterisation of individual topological transitions. As such, it provides a direct means of studying the means by which individual features are establishing connections with neighbouring features. The explicit definition in terms of the formalism of \emph{Persistent Topology}~\citep{edelsbrunner2010computational} provides us with a highly informative means of quantifying the intricate connectivity of the cosmic web, and hence represents the desired characterisation in terms of a solid mathematical foundation.

The key element of the present study concerns the change of the number of topological features as a function of spatial scale. As the scale changes, new features may emerge. This may involve the merging of formerly disconnected features, as well as the disappearance and annihilation of features. Examples of the latter happen when tunnels or cavities fill up. Such topological changes happen whenever the change in scale leads to the inclusion or removal of critical points, i.e., the minima, maxima or saddle points of the underlying manifold.

\subsubsection{Filtrations}
\label{subsec:filtrations}
To facilitate persistence analysis, for the analysis of the multi-scale nature of the mass distribution, the first step is a systematic definition of structural scale dependence. This leads to the concept of \emph{filtration}. It involves the quantification of structure in terms of a field $S$ that entails the characteristics of the mass distribution.

To establish the topological connections of the mass distribution, it is important to study the changes in topological structure as we proceed along different levels of the field $S$. This involves the emergence, merging, and disappearance of individual topological features, indicated as the \emph{birth} and \emph{death} of these features.

Mathematically, the set of filtrations of the field $S$ is the nested sequence of subsets $S_{j}$ for varying field levels:
\begin{equation}\label{eq:filtration}
 \emptyset = S_0 \subseteq S_1 \subseteq S_2 \subseteq ... \subseteq S_{M} = S. 
\end{equation}
Although the choice of filtration scale can allow for continuous values, a key observation is that we are only dealing with a finite number of critical points, and thus effectively with a finite number of filtrations. 

In a cosmological context, relevant filtrations involve a sequence of density field levels or a spectrum of spatial scales. In the case of the latter, for a discrete point sample we may accomplish this in terms of the (continuous) distance field\footnote{Note that there is a strong correlation between the distance field values and the density field: short distance field values correspond to regions of high density.}. Another option would be to circumvent the need for the calculation of the distance field, and seek to work directly from the object sample. It is the latter option which we pursue in this study.

\subsubsection{Density field filtrations}
In earlier works of our group ~\citep{pranav2017topology,pranav2019topology,pranav2019unexpected,feldbrugge2019,wilding2021}, we assessed the multi-scale topology of the cosmic mass distribution by means of the \emph{density filtrations}, where connections are formed based on the local density values. In this case, the field filtration is defined by the nested sequence of superlevel sets of the density field  $f(\textbf{r})$ by applying a threshold $\nu$ to the function values. The corresponding superlevel set $M(\nu)$ is
\begin{equation}
  \label{eq:superlevelsetDef}
  M(\nu)=\{(\textbf{r},f(\textbf{r})) \mid  f(\textbf{r})\geq \nu\},
\end{equation}
with the key \textit{nestedness property} 
\begin{equation}
  \label{eq:nesting}
  M(\nu_1)\subseteq M(\nu_2) ~~ \text{for $\nu_1 \geq \nu_2$}.  
\end{equation}
In view of this nestedness property, the superlevel sets $M(\nu)$ form a \textit{filtration}, i.e., a hierarchically ordered collection of nested sets as the density decreases from $+\infty$ to $0$.

\subsubsection{Distance field filtrations}
Another natural choice for the filtration of a field sampled by a discrete number of sample points is that of the distance field. It is defined as the value of the distance to a sample point, usually the closest sample point~\citep[see e.g.][]{aragon2010multiscale}. The focus on the distance field enables a more straightforward identification of features on a certain spatial scale: by defining the filtration of the field with respect to the distance values and grouping the sample points that get connected at a given distance threshold, we effectively identify structural features at the corresponding scale. For more detailed information on simplicial complexes, we refer to Appendix~\ref{app:simplicial}.

Hence, the nested sequence of the distance field filtration will result in a complete scale-space representation of the mass distribution~\citep{aragon2007,cautun2013nexus}. It will allow a detailed assessment of how structures at different spatial scales relate to each other and connect up in the cosmic mass distribution. 

Because the distance field is defined with respect to the discrete point distribution, we may actually not need the computation or reconstruction of a continuous distance field. Instead, we may work directly from the point distribution. To that end, the points are spatially connected by means of a \emph{Simplicial Complex} that reflects the density, geometry and multi-scale character of the spatial point distribution.

%%%%%%%%%%%
%%% BJ Moved to first citation of the figure
%%% Who knows what LaTeX will do with it?
%%% The first reference to this figure is just below here
%%%%%%%%%%%
\begin{figure*}
\centering
\includegraphics[width=0.9\textwidth]{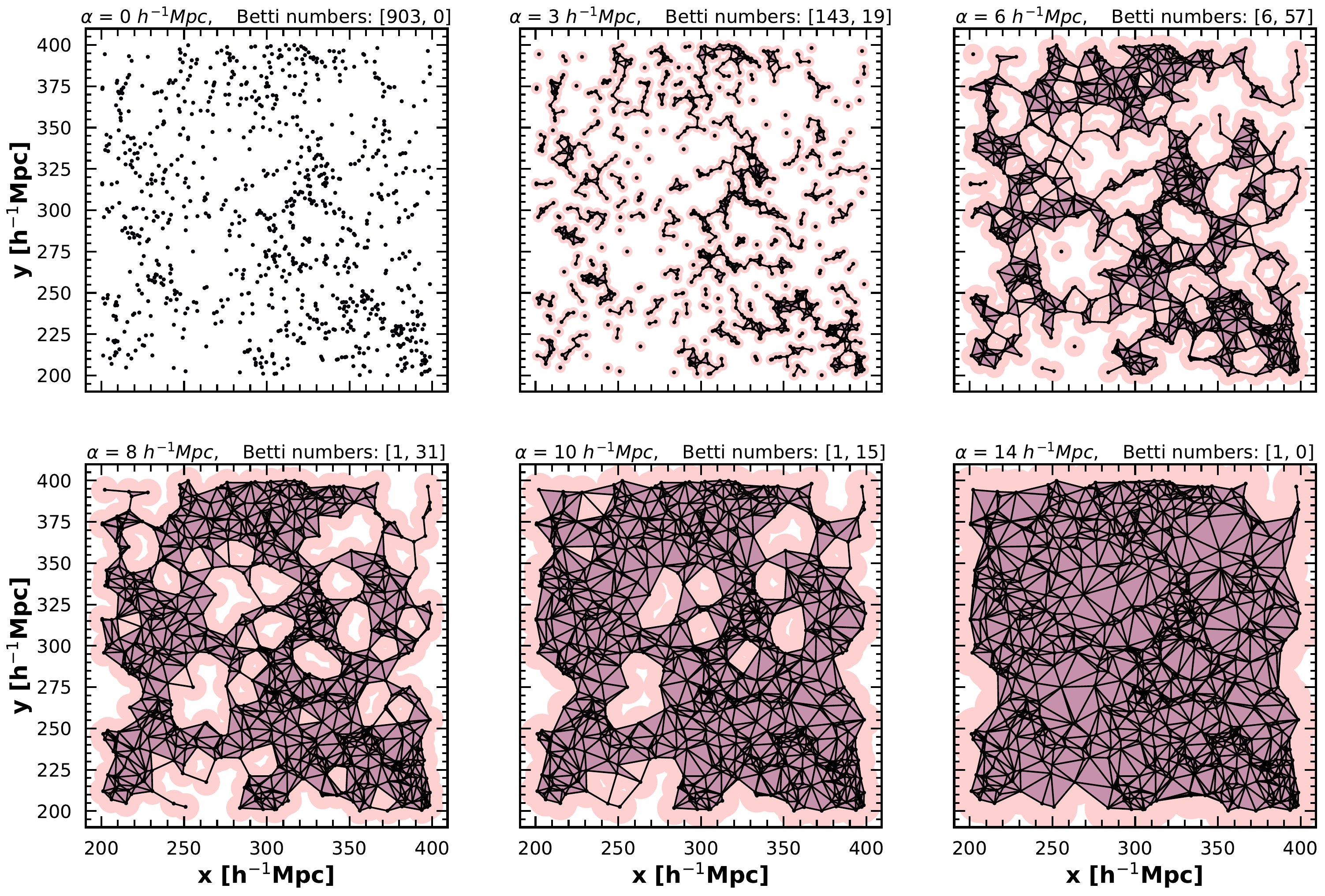}
\caption{\textbf{Alpha shape of a slice of haloes in the P-Millennium.} This plot represents the evolution of a two-dimensional alpha shape for a sample of haloes with a mass range of $\log_{10}\left(M\right) = \left(12.5,13\right] h^{-1}M_{\odot}$ in a slice of $30 h^{-1}$Mpc from the halo catalogue. Different panels are associated with different values of $\alpha$, representing different stages in the evolution of the alpha shape. From left to right in each row: $\alpha = 0 \ h^{-1}$Mpc, $\alpha = 3 \ h^{-1}$Mpc, $\alpha = 6 \ h^{-1}$Mpc, $\alpha = 8 \ h^{-1}$Mpc, $\alpha = 10 \ h^{-1}$Mpc, $\alpha = 14 \ h^{-1}$Mpc. 
These plots show how, as we increase the value $\alpha$, topological features (such as tunnels) are born and disappear and the haloes become more connected. The last panel represents an almost complete version of the Delaunay tessellation.}
\label{fig:1_AlphaShape_halos}
\end{figure*}

\subsection[Simplicial topology: Delaunay triangulations and alpha shapes]{Simplicial topology: \\ \ \ \ \ \ \ \ \ \ Delaunay triangulations and alpha shapes}
\label{sec:simplextop}
A simplicial complex is an ordered geometric assembly of faces, edges, nodes and cells that constitute a discrete spatial map of the volume containing a point set~\citep{edelsbrunner2010computational, pranav2017topology}. The geometric components of the complex are simplices of different dimensions: a cell is a three-dimensional simplex, a face or wall a two-dimensional simplex, an edge a one-dimensional simplex and a node a zero-dimensional simplex.

A good impression of a typical two-dimensional simplicial complex can be obtained from Fig.~\ref{fig:1_AlphaShape_halos}, showing a slice through a Delaunay tessellation (bottom right-hand panel) and a series of \emph{alpha shapes} (see Section~\ref{app:alphashape}). Zero-, one- and two-dimensional simplices are clearly visible: they correspond to the vertices and edges of the particle distribution's tessellation. 

The simplicial complexes at the center of the present study are Delaunay and Voronoi  tessellations~\citep{Voronoi1908,delone1934sphere}. 

%%%%%%%%%%%%
%%% From p7 Section 2.5 of manuscript 
%%% Moved to Appendix B as new subsection Delaunay triangulation}
%%%%%%%%%%%%

The self-adaptive nature of the Delaunay tessellations also assures us that it represents the topological structure traced by the sample points. Given the intention to assess the multi-scale topological aspects of the mass distribution probed by the sample points, it leads us to consider the possibility to define a scale based filtration of the Delaunay tessellation that is the simplicial representation of the probed mass distribution. The filtration of a simplicial complex $K$ is given by 
\begin{equation}
 \emptyset = K_0 \subseteq K_1 \subseteq K_2 \subseteq ... \subseteq K_m = K. 
\end{equation}
The hierarchical subsets $K_i$ of $K$ encode the change in topology throughout the simulation box.  It naturally leads to the concept of the scale-based filration of a Delaunay tessellation, known as \emph{alpha shapes}~\citep{Edelsbrunner1983,edelsbrunner1994,edelsbrunner2010computational}. Uniquely defined for a particular point set $P$ by a real number $\alpha$, the scale parameter, alpha shapes correspond to a unique assembly of simplicial (geometric) structures that capture the shape and morphology of the point distribution. They are a well-known concept in Computational Geometry and Computational Topology, and involve a generalization of the convex hull of a point set. For detailed information on and illustrations of alpha shapes see Appendix~\ref{app:alphashape}.

%%%%%%%%%%%%
%%% From p8 end of Section 2.5 of manuscript 
%%% Moved to Appendix B as new subsection Delaunay triangulation}
%%%%%%%%%%%%

\subsection{Persistent homology on alpha shapes}\label{subsec:persistenthomology}
\label{alphapersistence}
%%% -------------------------------------------------------------
Instead of studying the topology at a single length scale, alpha shape topology probes the topology of the distance field along its set of filtrations as a function of the length scale $\alpha$\footnote{Alpha shapes represent a mathematically well-defined filtration of the distance field, with a direct visual connection to the spatial pattern analyzed. There  are a range of other formalism and methods for the analysis of aspects of spatial point distributions that are based on the corresponding distance field. Amongst others, Minimal Spanning Trees and graph theoretical techniques are known for their application to the large-scale galaxy distribution. In Appendix~\ref{app:distancefield} we include a short discussion and comparison of several distance field based methods.}. As the filtration length scale $\alpha$ increases, the halo/sample point distribution proceeds through a process in which we see a growing number of Delaunay simplices getting connected as they get embedded in the corresponding alpha complex. As a result of this, we see a continuously changing population of different separate simplicial complexes and the corresponding structural entities they represent.

Proceeding through the entire value range of the scale parameter $\alpha$, we may follow the creation and destruction of the individual topological features. As new structures emerge, existing structures merge and other structures get annihilated, the number of (super)clusters of haloes, filaments and voids will continuously change. To obtain insight into the connection between the growing simplicial (alpha shape) complex and the topology of the spatial structure, it is illustrative to consider what happens as more points get added while the simplicial complex grows~\citep[see][]{weygaert2011alpha}. Cycling through the simplices of the alpha complex, we find the following:

When a point is added to the alpha complex, a new component is created: the zeroth Betti number $\beta_0$ is increased by 1.
In this study, the alpha complex starts out with all haloes added at $\alpha=0$, and each component corresponds to an individual halo. When adding edges between pairs of points, the components become (super)cluster of haloes. The process of adding edges between points can have two outcomes. Either both points belong to the same, or to different components of the current complex.
When they belong to the same component, the edge creates a new tunnel, increasing the one-dimensional Betti number by 1. The creation of a filament corresponds to the creation of a tunnel. When the edge connects two different components, two (super)clusters merge and $\beta_0$ decreases by 1. When a triangle gets added, it may complete the enclosure of a void or it may close a tunnel. In the first case, a new void is created and $\beta_2$ increases by 1. In the latter case, $beta_1$ is decreased by 1 as the number of tunnels and corresponding filaments decreases. Finally, when a tetrahedron is added, a void is filled. This means that the two-dimensional Betti number $\beta_2$ is lowered by 1.

By assessing the value of $\alpha$ at which a feature is created, the \emph{birth value}  $\alpha_b$, and the value at which a feature is destroyed, the \emph{death value} $\alpha_d$, we follow the existence of (super)cluster complexes, of filamentary features and of enclosed voids~\citep[see, e.g.,][]{wilding2021}. The difference between the creation and the destruction value,
\begin{equation}\label{eq:persistence}
    \pi \coloneqq \alpha_d-\alpha_b
\end{equation}
denotes the \emph{persistence} $\pi$ of a topological feature. By plotting the values of $\alpha_d$ vs. $\alpha_b$ for all topological features present in a spatial point distribution, for each dimensions $d=0,1,\ldots,D$, a highly detailed and idiosyncratic depiction of the topological structure is obtained. These \emph{persistence diagrams} not only quantify the topology in terms of a few summarising parameters, but in terms of diagrams that contain information on every individual topological feature present in the sample point distribution. 

Analogously, we can compute the Betti numbers ($\beta_0$, $\beta_1$ and $\beta_2$) of the distance field as a function of the length scale $\alpha$. The resulting Betti curves (i.e., the Betti numbers for all scales) provide an overview on the complete structure of critical points, and on how the global topology differs for varying scales. While Betti curves inform us of the overall topology (since they measure the total number number of topological features as a function of $\alpha$), persistence diagrams allow us to precisely track at what length scales each of these topological features is formed and for how long they \emph{persist}.

\subsection{Topological data analysis} \label{subsec:TDA}
%%% -------------------------------------------------------------
In order to obtain the persistence points and Betti numbers from the formalism described above, we used \textsc{Gudhi} (Geometric Understanding in Higher Dimensions). \textsc{Gudhi} is a generic open source C++ library for Computational Topology and Topological Data Analysis~\citep{GudhiGeneral}.

We start from the halo distribution in the  P-Millennium, sampled for a specific mass bin, and treating it as point-cloud data.  \textsc{Gudhi} allows the straightforward computation of the alpha complex~\citep{gudhi:AlphaComplex}, making use of the corresponding \textsc{CGAL} (the Computational Geometry Algorithms Library;~\citet{CGAL}) procedure. The simplices of the alpha complex are stored in a simplex tree~\citep{gudhi:FilteredComplexes}, each with the corresponding filtration $\alpha$. The persistent homology, i.e. the persistence points $(\alpha_i,\alpha_j)$, is then calculated using a persistent cohomology algorithm~\citep{gudhi:PersistentCohomology}. For a more detailed description of the algorithm, we refer to~\citet{GudhiHowTo} and~\citet{GudhiGeneral}. The Betti numbers that we will use to analyse the halo structure can be calculated directly from the persistence points. This is done for a chosen $\alpha$ by counting the pairs that were \textit{born} at a lower $\alpha$ and will \textit{die} at a higher $\alpha$.

\begin{figure*}
\centering
\includegraphics[width=0.8\textwidth]{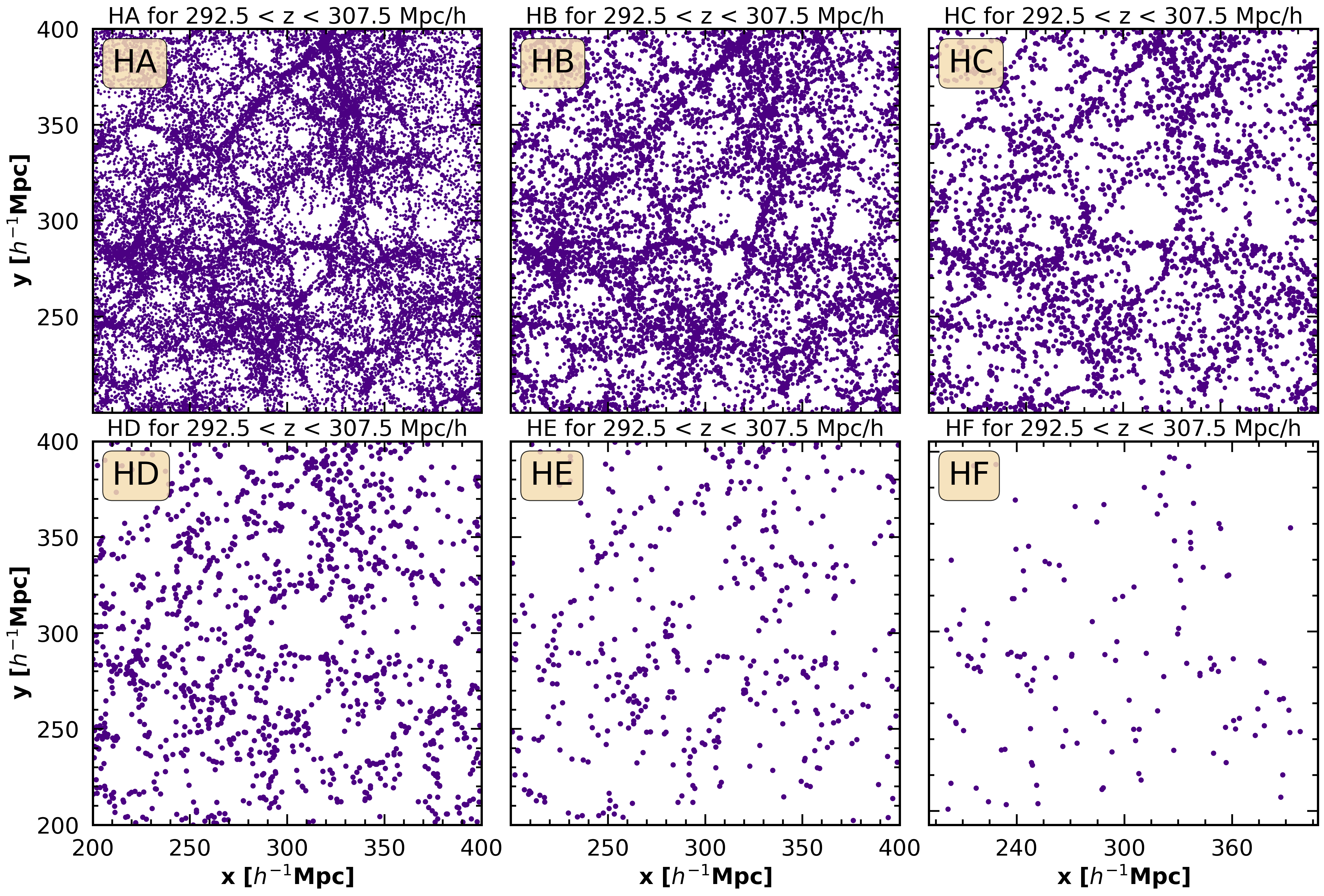}
\caption{\textbf{Spatial distribution of haloes in the P-Millennium.} This plot represents the point distribution of haloes of a $200\times200 \ h^{-1}$Mpc slice in the P-Millennium, cut through a slice of $15 \ h^{-1}$Mpc thickness. From left to right in each row: HA, HB, HC, HD, HE, HF. Notice that each of the panels represents the same simulation box, but with haloes subsampled from different mass ranges (see Table~\ref{tab:Classification Table} for the exact ranges of all halo populations). Each halo population highlights a different morphological aspect (or varying scales of connectivity) of the cosmic web, by sampling a different mass range. This is also reflected in the difference in number of haloes for each halo population, with the heaviest haloes tracing the high density nodes, and the lightest haloes weaving the whole filamentary structure. }\label{fig:2_RawSlicedDistribution}
%\end{figure*}
%\begin{figure*}
\centering
\includegraphics[width=0.8\textwidth]{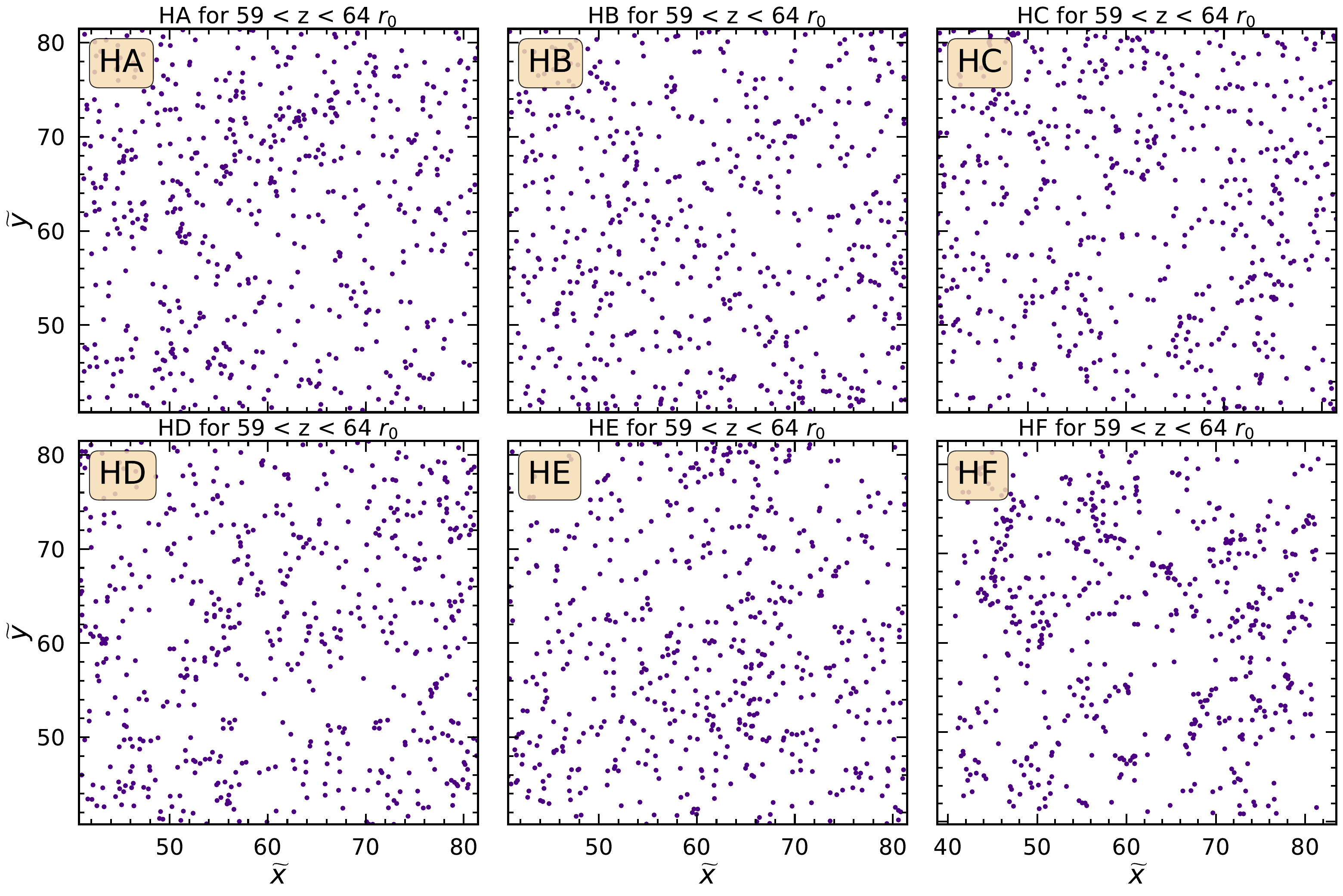}
\caption{\textbf{Re-scaled halo distribution in the P-Millennium.} The plot represents the point distribution of haloes of a scale-independent $80\times80$ ($r_0$) slice in the P-Millennium, cut through a scale-independent slice of $\Delta \widetilde{z} = 5$ thickness. From left to right in both rows: HA, HB. HC, HD, HE, HF. As in Fig.~\ref{fig:2_RawSlicedDistribution}, each of the panels represents the same simulation box, but with haloes subsampled from different mass ranges (see Table~\ref{tab:Classification Table} for the exact ranges of all halo populations) and their coordinates divided by each respective clustering length $r_0$ (see Table~\ref{tab:Clustering Table}). Additionally, the haloes of classes HA-HE have been randomly sampled to feature the same number of haloes within the slice as appear in HF.
We point out that whereas it is more difficult to distinguish between the different clustering behaviour of the mass classes, it is still possible. The heaviest haloes (HF) appear distinctly more clustered the classes of lighter haloes.}
\label{fig:3_ScaledSlicedDistribution}
\end{figure*}

% %########################################################
% %                         HALO POPULATION
% %########################################################

\section{Halo population} \label{sec:halopopulation}
\subsection{Simulation \& halo finding} \label{subsec: Halo Catalogue}
%%% -------------------------------------------------------------
In order to study the connectivity of haloes in the cosmic web, we use the state of the art $N$-body simulation for a standard $\Lambda$CDM model: the Planck-Millennium simulation (hereafter P-Millennium, see~\citet{Punya2018} and~\citet{Baugh2019}). The P-Millennium traces the cosmic evolution of over 128 billion ($5040^3$) dark matter particles of mass $M = 1.061\times10^8$ $h^{-1}M_{\odot}$ in a periodic box of sidelength $542.16 \ h^{-1}$Mpc. The P-Millennium compromises cosmological parameters from the latest Planck survey results~\citep{planck2018}: the Hubble-Lema\^itre parameter $h=0.6777$, where $h=H_0$/100 km s$^{-1}$ Mpc$^{-1}$ and $H_0$ is the Hubble-Lema\^itre constant. Density parameters correspond to $\Omega_{m}=0.307$, $\Omega_{\Lambda}=0.693$, and the density fluctuation is $\sigma_8 = 0.8288$. The evolution of these dark matter particles is traced back to $z=127$. However, in this work, we limit ourselves to the study of haloes in the  P-Millennium  at present time $z=0$.

The halo catalogue was obtained by running the halo and subhalo finder \textsc{Subfind}~\citep{Springel2001}. Subsequently, halo merger trees were obtained by using the \textsc{Dhalos} algorithm described in~\citet{Jiang2014}. A more detailed description of how the halo catalogue was obtained can be found in~\citet{Punya2018} and~\citet{Baugh2019}. We define the halo radius $R_{200}$ as the radius of a sphere enclosing a mass $M_{200}$. That is, each halo encloses an overdensity corresponding to $\Delta = 200$ times the critical density of the Universe. In total, there are approximately % $7.7\times10^7$ $3.5\times10^7$ of these dark matter haloes in the
P-Millennium. 

In this study we consider haloes more massive than $3.2\times10^{10} \ h^{-1}M_{\odot}$, corresponding to $1.13\times10^7$ of such haloes. This filtering has been chosen so that each halo is resolved by a large enough number of dark matter particles~\citep[300 particles, for more details see ][]{Punya2018, Bett2007}.
Nonetheless, our sample of $1.13\times10^7$ is large enough to characterise the topology of the cosmic web in a statistically representative manner. The haloes vary in mass from $\log_{10}\left(M\right) = \left[10.5, 13\right] \ h^{-1}M_{\odot}$. The halo catalogue
is into six mass bins, labelled as HA, HB, HC, HD, HE and HF. The mass ranges and number of haloes in these bins are shown
in Table~\ref{tab:Classification Table}.

\subsection{Mass dependence halo clustering}
Fig.~\ref{fig:2_RawSlicedDistribution} shows the distribution of a subsample of haloes of different mass ranges in the P-Millennium. As we consider less massive haloes, some morphological elements of the cosmic web such as filaments and voids appear more prominently. On the other hand, when considering the most massive haloes (HF), their assembly is distinctly different from their lower-mass counterparts. In the distribution of the most massive haloes (HF) the structures are larger and voids span greater distances than in the case of the least massive haloes (HA and HB). Simply by considering the spatial distribution of haloes in the P-Millenium, we observe how haloes of different mass ranges trace different structural patterns in the cosmic web. These structural patterns are precisely what our topological analysis aims to capture by means of the persistence diagrams and Betti curves.

The dependence of the spatial clustering of haloes on their mass has been known since the seminal study by~\citet{kaiser1987}. It led to the concept of \emph{bias} and the realisation that the spatial distribution of galaxies, haloes, and clusters that emerged from the cosmic matter field may offer a distorted view of the underlying dark matter structure. It manifests itself directly in terms of second order clustering, as measured by the two-point correlation function $\xi(r)$ of galaxies and haloes~\citep{peebles1980}. It involves the systematic trend of the correlation function amplitude with the mass of the objects. In general, biasing entails a higher clustering amplitude for more massive objects~\citep{kaiser1987,  Mo1996, Mo1997, Mo2002, MoBoschWhite2010, Desjacques2018}. The most outstanding case is that of the cluster distribution, where various studies have demonstrated the almost linear increasing trend of clustering amplitude with cluster mass and richness~\citep{Szalay1985, Bahcall1988, Bahcall2003, Berlind2006, Estrada2009}.

Based on this observation, we first address the two-point clustering of the different halo subsamples before investigating the
higher-order topological imprint. 

\begin{figure*}
\includegraphics[width=1\textwidth]{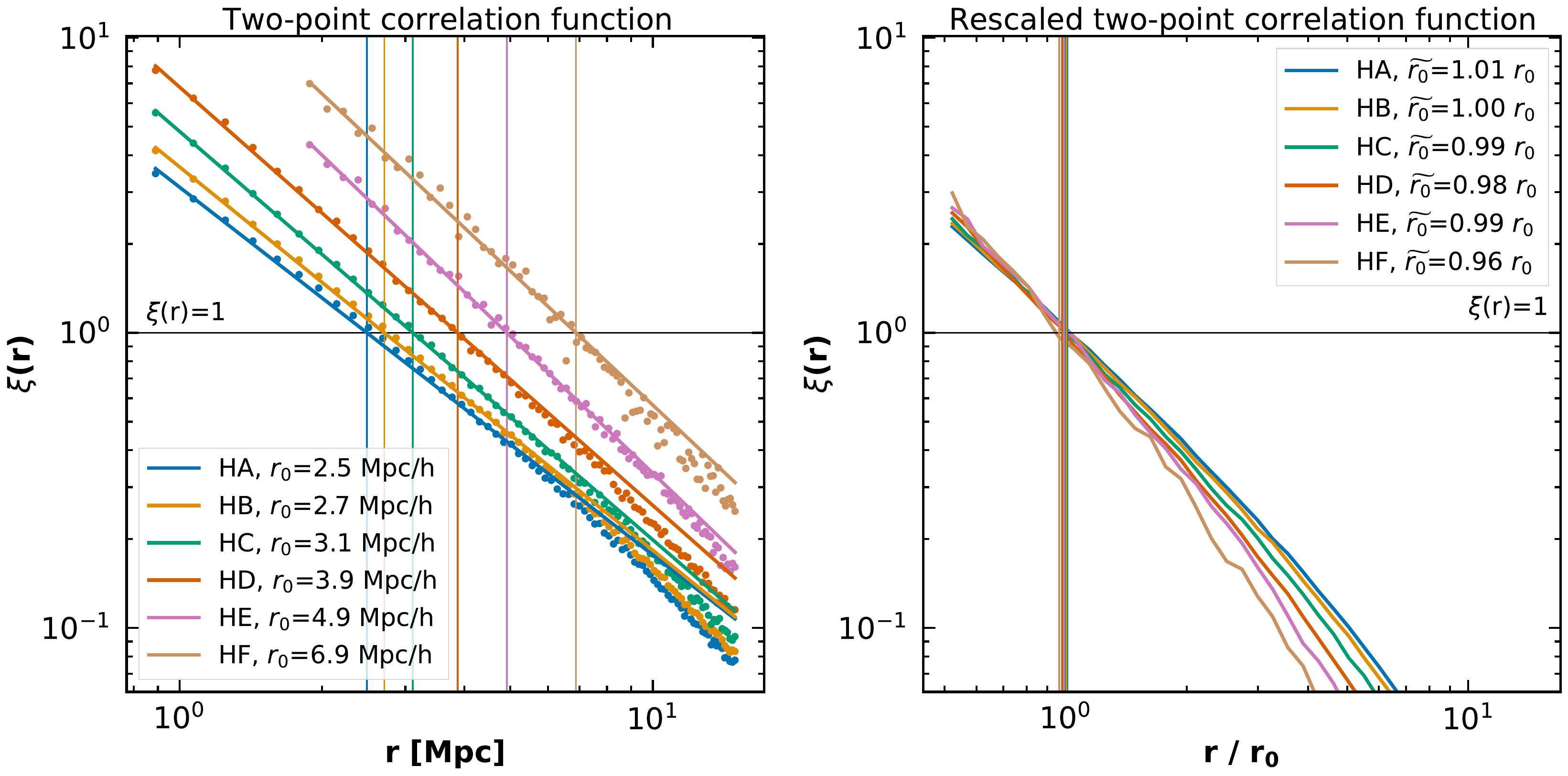}
\caption{\textbf{second-order clustering of haloes in the P-Millennium.} The plots show the two-point correlation functions for each of the halo populations of the P-Millennium. The left-hand panel shows the unscaled correlations functions, each with a powerlaw fit (see Table~\ref{tab:Clustering Table} for the parameters). This clustering information was used to rescale the halo locations, which is shown in the right-hand panel. The two-point correlation functions are shown as continuous lines, and the powerlaw fits in as dashed lines. For each panel we also show the (horizontal) line $\zeta(r)=1$ (in black), and vertical lines at each halo class's clustering length.  Where the clustering length $r_0$ different significantly in the unscaled case, rescaling leads to values very close to $\widetilde{r_0}=1$ (see Table~\ref{tab:Clustering Table} for the parameters). We observe a systematic trend where the massive haloes are more clustered than the lower-mass haloes.}\label{fig:4_tpcf}
\end{figure*}

\subsection{Clustering in P-Millenium subsamples}
%%% -------------------------------------------------------------
%%% BJ: moved correlation functin parameers of halo sub-populations here where it is first needed
%%% The main problem putting it here is that it already contains informatino about alpha-shape parameters.
%%% Original location is simply commented out so return to original location is easy
\begin{table*}
\begin{tabular}{@{}llllll@{}}
\hline
Halo Population & $\log_{10}$(M) \ {[}$h^{-1}M_{\odot}${]}  & $r_0$ \ {[}$h^{-1}Mpc${]} & $\gamma$  & $\widetilde{r_0}$ \ {[}$r_0${]} & $\widetilde{\gamma}$\\ \hline
HA & (10.5,11{]} & 2.49$\pm$0.01 & 	-1.247$\pm$0.008 & 	1.009$\pm$0.005 & 	-1.29$\pm$0.01 \\
HB & (11,11.5{]} & 2.71$\pm$0.01 & 	-1.298$\pm$0.006 & 	1.003$\pm$0.004 & 	-1.33$\pm$0.01 \\
HC & (11.5,12{]} & 3.11$\pm$0.01 & 	-1.385$\pm$0.004 & 	0.995$\pm$0.003 & 	-1.41$\pm$0.01 \\
HD & (12,12.5{]} & 3.87$\pm$0.02 & 	-1.417$\pm$0.008 & 	0.979$\pm$0.003 & 	-1.50$\pm$0.01 \\
HE & (12.5,13{]} & 4.91$\pm$0.02 & 	-1.54$\pm$0.01 	& 	0.99$\pm$0.01 	& 	-1.55$\pm$0.02 \\
HF & (13,13.5{]} & 6.88$\pm$0.08 & 	-1.51$\pm$0.02 	& 	0.963$\pm$0.007 & 	-1.72$\pm$0.03 \\ \hline

\end{tabular}
\caption{\textbf{Clustering correlation parameters of haloes in the P-Millennium simulation}. For each halo class and its logarithmic mass range we show the respective clustering length $r_0$ and the exponent of the power law fit $\gamma$. $\widetilde{r_0}$ and $\widetilde{\gamma}$ are the same parameters for the halo populations with second-order clustering normalised using the respective clustering lengths $r_0$.}
\label{tab:Clustering Table}
\end{table*}

In our P-Millenium halo sample, the mass dependence of halo two-point correlation function is clearly borne out by its behaviour for the
different halo mass bins (see left-hand panel of Fig.~\ref{fig:4_tpcf}). To a first approximation, the two-point correlation function behaves as a power-law,
\begin{equation}
\label{eqn: tpcf}
    \xi\left(r\right) = \left(\frac{r}{r_0}\right)^{\gamma}\,,
\end{equation}
so that it is fully characterized in terms of the power law slope $\gamma$ and the
clustering amplitude in terms of its correlation or clustering length $r_0$. Table \ref{tab:Clustering Table} (3rd and 4th column) lists the
parameters $r_0$ and $\gamma$ for the P-Millenium halo subsamples. \footnote{Note: to compute the two-point correlation function of each halo population in the P-Millennium, we use the \emph{Landy-Szalay} estimator~\citep{LandySzalay1993,Kerscher2000}, which is given by 
\begin{equation}
\label{eqn: Landy Estimator}
    \xi\left(r\right) = \frac{\langle DD \rangle - \langle 2DR \rangle + \langle RR \rangle}{\langle RR \rangle},
\end{equation}
where $\langle DR \rangle$ is the normalised number of pair counts between the data and the Poisson point distribution while $\langle DD \rangle$  is the normalised number of pair counts between the data. Lastly, $\langle RR \rangle$ corresponds to the normalised number of pair counts between the Poisson point distribution. }. There is a clear systematic trend, where more massive haloes (such as HD, HE and HF) are more correlated and clustered than the lower mass haloes (HA, HB and HC).

\subsection{Clustering Rescaling} \label{subsec: rescaleclustering}
%%% -------------------------------------------------------------
To get a more balanced and objective impression of the higher order clustering aspects 
of the halo (sub)samples  we compensate for two aspects of the halo distribution, the
differences in clustering strength and the differences in number density of the
halo subsamples.

We compensate for the stronger clustering of the more massive halo samples by rescaling the various
subsamples by the corresponding clustering scales $r_0$. The scales, and coordinates $x,y,z$
of the haloes, are rescaled to dimensionless coordinates 
\begin{equation}
         \widetilde{x} \coloneqq \frac{x}{r_0}, \qquad
         \widetilde{y} \coloneqq \frac{y}{r_0}, \qquad
         \widetilde{z} \coloneqq \frac{z}{r_0} \, .
    \label{eqn: Re-scaled coordinates}
\end{equation}
The two-point correlation function of the rescaled halo samples is plotted in Fig.~\ref{fig:4_tpcf} (righthand panel). As intended,
the rescaled correlation function have equal clustering length $\widetilde{r_o}=1$. They are also similar power-law functions,
with almost equal slopes at small scales $\widetilde{r}<2$. However, towards larger scales the rescaled correlation functions
do clearly reveal systematic differences, with the higher mass samples displaying a stronger decline towards larger scales.
Apparently, the different halo subsamples are not perfect selfsimilar representatives of an underlying distribution. Instead,
this testifies of the significant presence of higher order clustering contributions. 

Visually, the differences in clustering patterns between the halo (sub) samples reveal themselves optimaly by sampling
the same number of haloes within a given rescaled volume. This removes the influence of point density on the
impression of clumping perceived by the human eye. Fig.~\ref{fig:3_ScaledSlicedDistribution} plots the rescaled
spatial distribution of haloes in the six subsamples in the same rescaled volume, a slice of (rescaled) size $80 \times 80 \times 5$.
Each of the halo samples contains a random selection of the same number of 200 haloes. 
 
The overall impression is that of spatial patterns that largely resemble each other. This is a clear manifestation of the
approximately selfsimilar character of the halo distribution as expressed by the power-law two-point correlation function.
Overall, even the large scale structure of the different halo distribution is largely similar. Nonetheless, we also discern a gradual
and systematic shift in the nature of the large scale patterns defined by the halo distributions, from a somewhat random character
for the HA sample (top lefthand panel) towards a more and more structured geometric pattern in the most massive halo sample HF
(bottom righthand panel). 

The subtle differences in structure at large scales, also in the rescaled halo distributions, are responsible for the
differences seen in the corresponding rescaled second-order correlation functions and deviations from pure selfsimilarity.
They testify of the presence of higher-order correlations. Here, we therefore aim to identify structural patterns that go beyond
the second-order probe of clustering. The challenge is to identify and quantify this in terms of measures that facilitate a
direct relation with the visible weblike spatial pattern, its multiscale character and connectivity of structure in the
distribution of haloes. It leads us to the investigation of the nature of these higher-level structures with a topological
underpinning, and a description in terms of \textit{persistent topology}.

%########################################################
%                         RESULTS
%########################################################
\section{Topological scale dependence} \label{sec: Results}
In this section, we delve into the scale-dependence of the halo topology. First, we will treat the overall topology and connectivity of the cosmic web, traced through the different samplings of the six halo mass classes. This is done by looking at the total number of topological features (for each dimension) at different scales, represented by the Betti curves for varying filtration length scales
$\alpha$. This allows a visualisation of the global build-up of the connectivity, highlighting general characteristic length scales.
We then focus on the much more precise characterisation of the length scales $\left(\alpha_b,\alpha_d\right)$ at which each of the topological features form and disappear, provided through the persistence diagrams. This will give insights into the multi-scale nature of the cosmic web as traced by dark matter haloes.

Both Betti curves and persistence diagrams serve as topological tracers of the strength of the correlation between the structures haloes outline and the underlying geometric dimension. These tracers (after normalising according to the two-point correlation function) quantify higher orders of structuring and the correlation of the halo mass with the underlying geometry. As such they pick up different clustering behaviour of light and heavy haloes, and allow statements about how it depends on the geometrically defined environment.

\subsection{Alpha Shapes of the halo distribution}
Fig.~\ref{fig:1_AlphaShape_halos} shows the two-dimensional alpha shape for the HC halo population inside a thin slice through
the simulation volume. In a sequence of six panels, we see the development of alpha shapes for increasing value of its scale
parameter $\alpha$. The figure highlights the sensitivity of the alpha shape to the clustering of haloes and its multiscale
nature, revealing the presence and scale of filamentary bridges, gaps or voids of haloes, and tunnels that only fill up at the
highest values of $alpha$. 

The vast potential and richness of the topological information on the cosmic web contained in alpha complexes becomes even more 
apparent when assessing the full three-dimensional situation. Figs.~\ref{fig:B1_3d_alphashape} and \ref{fig:B2_3d_alphashape_zoom}, 
Appendix~\ref{app:alphashape}, testify of this. They show -- for three different values of $\alpha$ -- the full three-dimensional
alpha shapes for the HA halo population of the $P$-Millennium simulation, in a slice with the full simulation box width of
$542.16 \ h^{-1}$Mpc and 30 $h^{-1}$Mpc width, along with zoom-ins into specific regions. They reveal in meticulous detail the
weblike pattern defined by the distribution of haloes, the nature of its composing structures, their mutual connections and
its intricate multiscale character. Further details are outlined in the corresponding Appendix~\ref{app:alphashape}. 

%%% BJ: We moved this from the prevoius section since this is about rescaling
%%% The original is left in place commented for recovery if needed.

\subsection{Scale-independent parametrization -- the sample rescaling} \label{subsec: Re-scaling}
%%% -------------------------------------------------------------
Following our observation in the previous section that the higher order clustering character of the halo distribution manifests
itself directly in the pattern of the corresponding \textit{rescaled} spatial point distribution, we assess its multiscale
topology by analyzing the \textit{rescaled alpha shapes}. This entails the rescaling of the scale parameter $\alpha$ by the
corresponding clustering length $r_0$, 
\begin{equation}
    \widetilde{\alpha} \coloneqq \frac{\alpha}{r_0},
\label{eqn: Re-scaled alpha}
\end{equation}
and with respect to the topological persistence diagrams the rescaling of the birth and death scales $\widetilde{\alpha}_b$
and $\widetilde{\alpha}_b$ of the various topological features,
\begin{equation}
    \widetilde{\alpha}_b \coloneqq \frac{\alpha_b}{r_0}, \qquad
    \widetilde{\alpha}_d \coloneqq \frac{\alpha_d}{r_0}\,. \qquad
\label{eqn: Re-scaled alphabd}
\end{equation}

In the case of a purely self-similar spatial pattern, fully specified by its two-point correlation function, we
would expect its topological properties to remain unaltered for the halo subpopulations. The topology would be entirely
determined by the second order clustering. However, in the presence of higher order clustering the situation will be
different resulting in more complex topological behaviour. The different spatial patterns seen in the scaled
halo distributions in different mass ranges testify of such topological bias. To optimize the sensitivity to
these topological properties, we therefore assess the Betti curves and persistence diagrams for both the
regular and the scaled halo distributions. 

\begin{figure*}
   \includegraphics[width=0.8\textwidth]{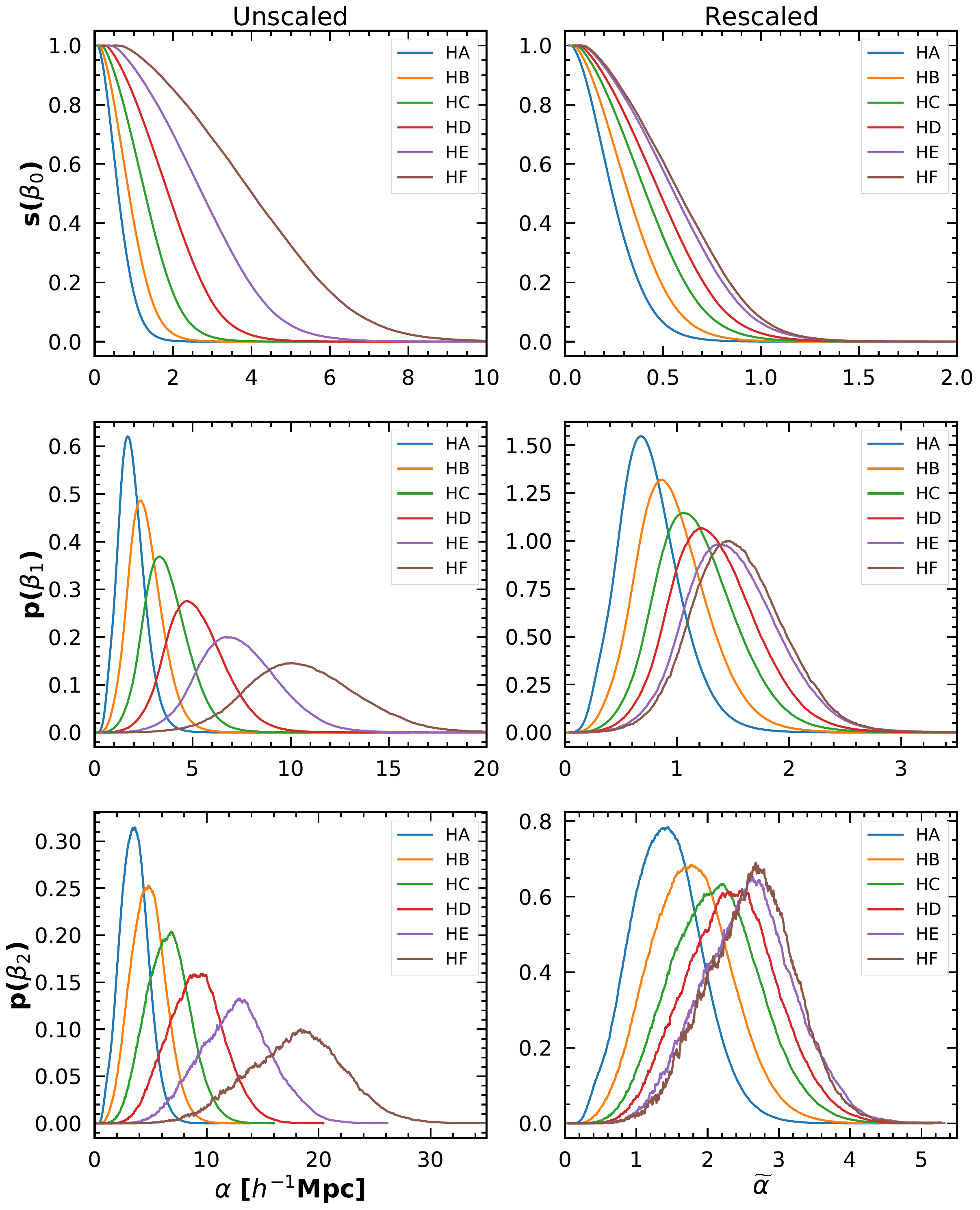}
\caption{\textbf{Betti curves for different classification of haloes in the P-Millennium.} From top to bottom: standardised $\beta_0$-curve $s\left(\beta_0\right)$, corresponding to disconnected haloes, normalised $\beta_1$-curve $p\left(\beta_1\right)$, associated with filaments and $\beta_2$-curve $p\left(\beta_2\right)$, associated with voids. The left column corresponds to the unscaled Betti curves, where $\alpha$ is in physical units of $h^{-1}$Mpc, whereas the right column corresponds to the re-scaled Betti curves, where $\widetilde{\alpha} = \frac{\alpha}{r_0}$ represents a scale-independent parameter that nullifies the clustering of the two-point correlation function (see Table~\ref{tab:Clustering Table} for exact values of $r_0$).
Note that even when we take out the two-point correlation function from the Betti curves (by considering the evolution of the alpha shape in terms of the scale-independent parameter $\widetilde{\alpha}$), the haloes still present a topology that is not self-similar with respect to each other.}\label{fig:5_BettiCurves}
\end{figure*}

As we will observe in the following section, also the scaled Betti curves and persistence diagrams reveal a systematic
shift from low mass to high mass haloes. The systematic shift shows that different halo populations (of different mass ranges) trace different structural patterns at higher-order clustering, and hence reflect a topological bias.

% -------------------------- BETTI CURVES --------------------------------------
\subsection{The overall topology -- Betti curves} \label{subsec: Results Betti curves}
%%% -------------------------------------------------------------

We first investigate the global topology by analysing the Betti curves for the halo distribution in the P-Millennium. We show the number of independent topological features per dimension $(\beta_0, \beta_1, \beta_2)$ for both unscaled (such that $\alpha$ represents a physical distance) and re-scaled (with respect to $r_0$) cases to highlight the general multi-scale buildup. To ease the visualisation of these curves, the $\beta_0$-curves were standardised by dividing by the number of haloes in each halo population, indicated as $s\left(\beta_0\right)$. In this case, at a value of $\alpha=0$, the disconnected haloes corresponds to $s\left(\beta_0\right) = 1$, as no connections have taken place in the alpha shape yet and the number of separate clusters of haloes is equal to the total number of haloes.

For the $\beta_1$- and $\beta_2$-curves (associated with filamentary loops and voids respectively), we carried out a normalisation of the Betti curves (dividing by the area under the curve), which we denote as $p\left(\beta_i\right)$ for $i \in \{1,2\}$. Notice that these procedures were applied for both unscaled and re-scaled Betti curves, as shown in Fig.~\ref{fig:5_BettiCurves}.

\subsubsection*{$\beta_0:$ Disconnected haloes}
%%% '''''''''''''''''''''''''''''''''''''''''''''''''''''''''''''
The first row of Fig.~\ref{fig:5_BettiCurves} shows the $\beta_0$-curves. For the unscaled curves (top left panel), we observe a consistent trend. That is, the lower-mass haloes connect up at lower values of $\alpha$ as the number of them (per population) is higher. More precisely, the lowest-mass haloes (HA) have all connected at length scale of $\alpha = 2 \ h^{-1}$Mpc whereas the most massive haloes are still not fully connected at length scales of $\alpha = 9 \ h^{-1}$Mpc.
This indicates that it is more likely for lower mass haloes to find another such halo of the same mass range within their neighbourhood and that higher mass haloes are spread out over larger structures with connections forming for larger length scales.

As soon as we correct the clustering length of each halo population with respect to $\alpha$ interesting trends start to appear. The $\beta_0$-curves start to become more similar with respect to each other in terms of their steepness and the speed with which they attain full connectivity. Since the distance at which haloes form connections is associated with their clustering length, re-scaling the $\beta_0$-curves with $r_0$ provides a more comparable description of the topology of the different halo populations. 
More explicitly, we see now that the lowest-mass haloes have mostly connected at a scale-independent threshold of approximately $\widetilde{\alpha} = 0.5$, while the most massive haloes have formed most connections only at $\widetilde{\alpha} = 1.5$. This is concrete indication that the heaviest haloes trace features on larger scales than their lighter counterparts, with some relevant connections still forming even at re-scaled distances exceeding the clustering length as determined from the second-order clustering. We point out this behaviour in Fig.~\ref{fig:2_RawSlicedDistribution} (unscaled) and Fig.~\ref{fig:3_ScaledSlicedDistribution} (re-scaled).

\subsubsection*{$\beta_1:$ Filaments}
%%% '''''''''''''''''''''''''''''''''''''''''''''''''''''''''''''
The second row of Fig.~\ref{fig:5_BettiCurves} shows the $\beta_1$-curves. Again, we see a consistent trend for the unscaled $\beta_1$-curves (middle-left panel). The lower-mass haloes (HA-HC) form most of their filamentary structure at values of $\alpha < 5  \ h^{-1}$Mpc.
In the case of the most massive haloes (HE, HF), we see that most filaments appear at larger scales of $\alpha >  \ 7.5 \ h^{-1}$Mpc. This indicates the birth of large-scale filaments and a highly connected (as indicated by the behaviour of the $\beta_0$-curves at the respective values of $\alpha$) filamentary structure that makes up the complex network of the cosmic web. The multi-scale nature of these structures is better apparent from the persistence diagrams, whereas the Betti curves serve as a straight-forward indication of overall trends and large-scale changes in the halo-topology.

Even when we correct for the clustering length of each halo population (middle-right panel), we still observe the same trend as for the unscaled case. More massive haloes (HD-HF) form filamentary loops later than the less massive ones (HA-HC). The majority of filamentary loops are formed at values of $\widetilde{\alpha} < 1$ for the three lowest mass classes, and for values between $1.5 < \widetilde{\alpha} < 2.5$ for the three classes of heavier haloes. These differences together with the broader curves for the heavy haloes (HD-HF) indicates that more massive haloes are not merely tracing a scaled-up version of the structures traced by lighter haloes, but that their connected structures form and exist over wider ranges in the probed distance space.

\subsubsection*{$\beta_2:$ Voids}
%%% '''''''''''''''''''''''''''''''''''''''''''''''''''''''''''''
The last row of Fig.~\ref{fig:5_BettiCurves} shows the $\beta_2$-curves. Similarly to the intensity persistence diagrams of dimension 2, the bottom left panel indicates that lower-mass haloes (HA-HC) form most of their voids at length scales of $\alpha<8 \ h^{-1}$Mpc. More remarkable is the case of the more-massive haloes (HD-HF). For these halo populations, the $\beta_2$-curves manage to capture a rich hierarchy of void formation at length scales of $\alpha > 10 \ h^{-1}$Mpc. 

\begin{figure*}
\centering
\includegraphics[width=0.85\textwidth]{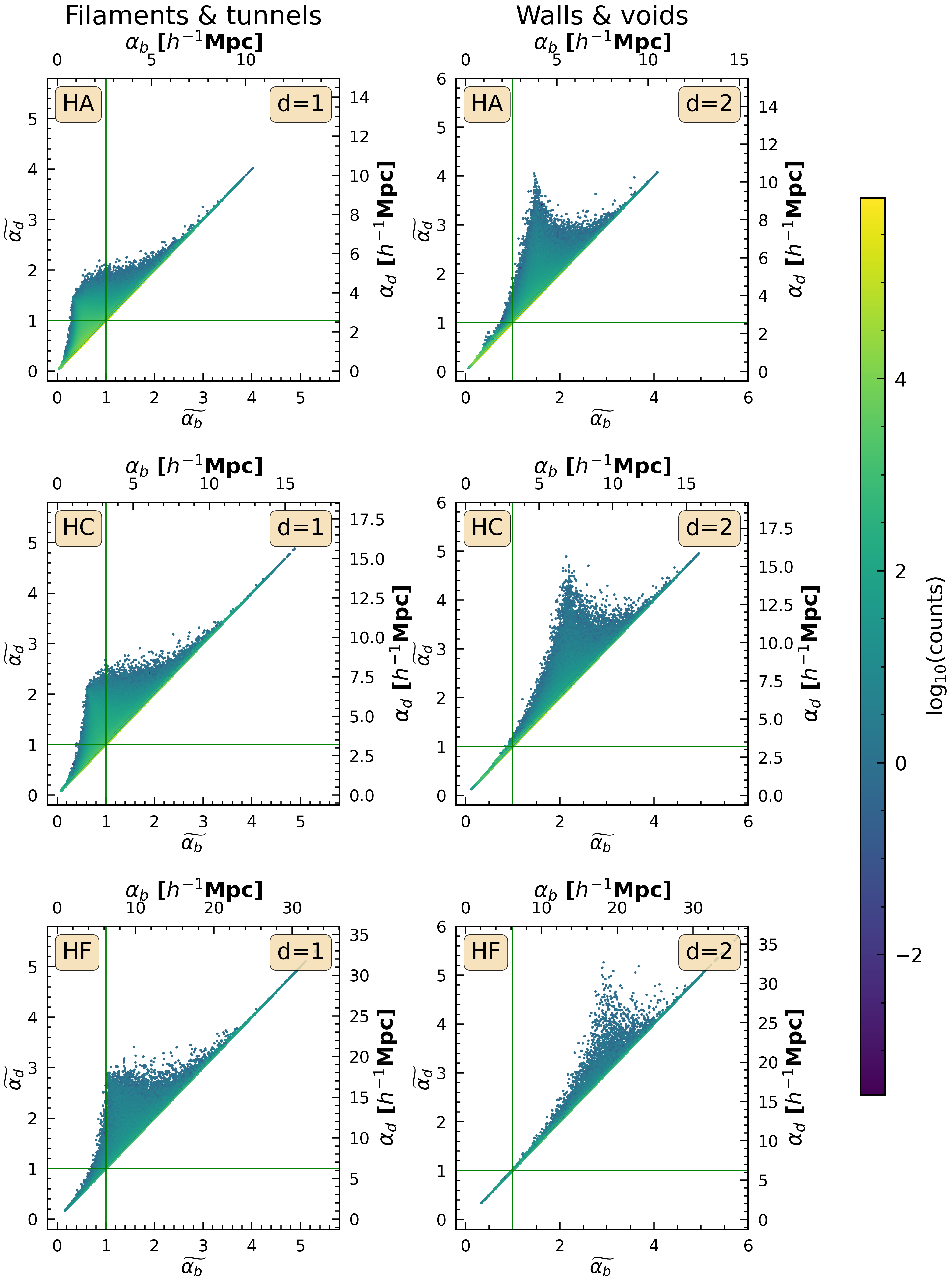}
\caption{\textbf{Persistence diagrams.} We show the diagrams of dimension one (associated with filaments, left column) and dimension two (associated with voids, right column) for three different classifications of haloes in the P-Millennium. From top to bottom we show the diagrams of haloes of the classes HA, HC and HF (see Table~\ref{tab:Clustering Table} for the exact mass ranges). The birth- and death-thresholds ($\alpha_b, \alpha_d$ respectively) are given both in units of $h^{-1}$Mpc (secondary axes) and in the scale-independent parameters $\widetilde{\alpha}_b$ and $\widetilde{\alpha}_d$ (primary axes). The manifest differences in these birth-death diagrams between the two columns reveal the differences between haloes of a given mass in either of the two topologically defined environments. This is clear evidence for a topological bias.}\label{fig:6_intensity_persistence}
\end{figure*}

Although this hierarchy becomes more diffuse when we correct for the clustering length of each halo population, we still observe a difference in the formation of voids with respect to $\widetilde{\alpha}$. We observe a similar symmetry to that of filament formation around $\widetilde{\alpha}= 1$. The void structure of lower-mass haloes is present at scale-independent values of $\widetilde{\alpha} \leq 2$, whereas the most significant voids only appear at scale-independent values of $\widetilde{\alpha} > 2$, corresponding to the more massive haloes.

% -------------------------- PERSISTENCE DIAGRAMS --------------------------------------
\subsection{Multi-scale connectivity -- persistence diagrams} \label{subsec: Results IPDs}
%%% -------------------------------------------------------------
Following the formalism presented in Section~\ref{subsec:TDA} and Appendix~\ref{app:alphashape}, we computed the persistence points $\left(\alpha_b,\alpha_d\right)$ and plotted $\alpha_d$ vs. $\alpha_b$. Given the large number haloes in the  P-Millennium, we computed what we denote as \emph{intensity persistence diagrams}~\citep{pranav2017topology}. These are equivalent to the standard persistence diagrams, except that a colour map indicates the number of counts of persistence points of a certain region on the persistence plane. This was done by constructing a $2D$ histogram counting, for every persistence point, the number of other pairs located in the histogram bin.

In Fig.~\ref{fig:6_intensity_persistence} we depict the one- and two-dimensional intensity persistence diagrams of three halo populations (HA, HC and HF). The zero-dimensional diagrams are not shown and excluded from this discussion, as all features share a birth-value $\alpha_b$ of zero, and they are thus less informative than in the other dimensions. We are mainly interested in studying the characterisation of filaments (i.e. filamentary loops, associated with one-dimensional topological features) and voids (associated with two-dimensional topological features). The primary axes of the diagrams (bottom and left) use the re-scaled values $\widetilde{\alpha}$, whereas the secondary axes (top and right) show unscaled values $\alpha$. The cluster length ($\widetilde{\alpha}_{b,d} = 1$) is marked by horizontal and vertical lines. In these dimensions the persistence diagrams exhibit a characteristic, roughly triangular shape, that can be attributed to the multi-scale nature of the structures under investigation.

To a large degree, their shape reflects what can be expected from a halo-based sampling of the underlying dark matter distribution. The detailed information on the process of structure formation due to gravity that can be obtained from the persistence diagram of the dark matter distribution is discussed in an earlier publication of the group~\citep{wilding2021}. Several of the relevant notions also extend to the persistence diagrams of the halo distribution, whereas there are also notable deviations.

%%% BJ: Just so the referee can find it we makea subsubsection.
\subsubsection*{The Apex of the persistence diagram}
%%% '''''''''''''''''''''''''''''''''''''''''''''''''''''''''''''
Fig.~\ref{fig:6_intensity_persistence} illustrates the Intensity persistence diagrams.  The roughly triangular shape of the diagrams is the first similar characteristic. In general, the triangular region is bounded by the diagonal and two edges converging towards a tip. For the tip of this region in particular, we coined the term \emph{apex} of the persistence diagram~\citep{wilding2021}. At least one of the edges commonly exhibits a concave shape, in the case that both edges feature this attribute the diagram usually spouts a distinct, sharp apex.

The base of the triangular region is formed by persistence points along the diagonal. These points, with very small values of persistence, are the most short-lived topological features commonly associated with topological noise. Opposed to that, the apex consists of the points with the highest values of persistence. These points characterise the most prominent features of the halo distribution -- they are visible during wide ranges of $\alpha$.

The apex and its sharpness also indicate the occurrence of phase transition-like behaviours in the emerging connectivity of the cosmic web. Such a behaviour is generally found in regions where a large number of persistence points appear with similar birth- or death-values ($\alpha_b$ or $\alpha_d$). An example of such a transition can be found in the two-dimensional persistence diagram of the lightest haloes (HA), where the apex is particularly tapered. The features in the apex indicate prominent structures that emerge within a very narrow range of birth-values $\alpha_b$.
The location of the apex (in terms of unscaled/re-scaled birth- and death-values) will also serve as a tracer for the most prominent and distinct structures. The main variations between in the persistence diagrams shown here occur in relation to the shape and location of the apex.

\subsubsection*{Dimension 1: Filaments}
%%% '''''''''''''''''''''''''''''''''''''''''''''''''''''''''''''

In the left column of Fig.~\ref{fig:6_intensity_persistence} we show the one-dimensional intensity persistence diagrams for the halo populations HA, HC and HF. The diagrams outline the mutli-scale filamentary structure traced by the dark matter haloes, separate for each population. From the diagram it is possible to analyse the emergence (birth), disappearance (death), scale, and prominence of topological features, which in this case are independent, closed filamentary loops.
Immediately apparent is the flattened or rounded apex of the diagram for the lightest haloes (HA), which makes it difficult to pinpoint exact associated birth- or  death-values. For the lightest haloes the most persistent features are born in a small range slightly below $\widetilde{\alpha}_b \approx 0.5$ and die at values below $\widetilde{\alpha}_d \approx 2.0$.

Moving to higher mass haloes leads to a clear sharpening of the apex, and the concentration of points with high persistence in more narrow ranges of $\widetilde{\alpha}_b$ and $\widetilde{\alpha}_d$. For the HC halo population, the slightly more narrow apex is located at birth-values in the range of $\widetilde{\alpha}_b \simeq 0.5$ to $1.0$, with death-values around $\widetilde{\alpha}_b \simeq 2.3$.

For the heaviest haloes the characteristic length scales of the prominent filamentary features shifts to even higher densities, with the apex now converging into a sharp tip. This occurs at a birth-value of $\widetilde{\alpha}_b = 1$, and at a death-value of slightly below $\widetilde{\alpha}_d = 3$.

The changes in the shape of the apex are a clear indication that when considering a higher mass class of haloes, on the one hand prominent features traced by those haloes form connections at more similar length scales. On the other hand, the number of structures of comparable persistence (i.e. with similar distance to the diagonal) decreases, going from a larger number of features (in a flat, plateau-like apex) forming well below the respective clustering length (HA) to much fewer features forming at almost exactly the clustering length (HF).

Naturally, very massive haloes reside in massive filaments. But the clustering length, as a characteristic probe of the degree of clustering (and thus mass) of a halo population, together with its relation to the existence and location of topological features of high persistence (i.e. massive loops of filaments) allows deeper insights. In particular the differences in the location of high-persistence features when the halo mass changes allows the identification of higher order trends between halo mass and the underlying filamentary network.

\subsubsection*{Dimension 2: Voids}
%%% '''''''''''''''''''''''''''''''''''''''''''''''''''''''''''''
The right-hand column of Fig.~\ref{fig:6_intensity_persistence} shows the two-dimensional intensity persistence diagrams for the halo populations HA, HC and HF. These diagrams trace the multi-scale nature of the void structure as woven by dark matter haloes.

Whereas the one-dimensional diagrams (tracing the filamentary components of the halo structure) shown in the left-hand column feature three very differently shaped apexes, the shapes of the two-dimensional diagrams' apexes are largely similar to each other. They differ mostly in their exact location, and in their prominence, determined by the number of points they consist of, which is mainly due to the significantly varying numbers of haloes in the different mass classes (see Table~\ref{tab:Classification Table}).

The apexes exhibit a clear shift in location, clearer for the one-dimensional diagrams. Regarding the lowest-mass haloes (HA), we observe that the most persisting voids are formed at birth-values of $\widetilde{\alpha}_b = 1.5$ and filled at $\widetilde{\alpha}_b = 3$ to $4$. Interestingly, we observe the appearance of a small peak in persistence at values of $(\widetilde{\alpha}_b,\widetilde{\alpha}_d)\simeq (0.75,0.75)$. This likely corresponds to a small collections of low mass haloes connecting up to form local voids at very small scales. However, these do not represent the cosmic vast voids that we are interested in quantifying in this study.

Moving to more massive haloes (HC), we see that the apex has shifted to higher birth- and death-values. Most of the void formation now occurs at a birth-value of $\widetilde{\alpha}_b\simeq2$, and most of these voids are filled at values between $\widetilde{\alpha}_d=4$ to $5$.

For the case of the most massive haloes (HF), most of the void formation occurs at a higher birth-threshold of $\widetilde{\alpha}_b \simeq 3$. Moreover, these large voids persist to death-thresholds of $\widetilde{\alpha}_d \simeq 5.25$. These topological features correspond to voids of persistence values of approximately $\pi \approx 15 \ h^{-1}$Mpc, and indeed seem to be in accordance to the spatial distribution of haloes as shown in Fig.~\ref{fig:2_RawSlicedDistribution}.

% -------------------------- LIFETIME CURVES --------------------------------------
\subsection{Distribution of persistence -- lifetime curves}
%%% -------------------------------------------------------------

\begin{figure*}
   \includegraphics[width=0.9\textwidth]{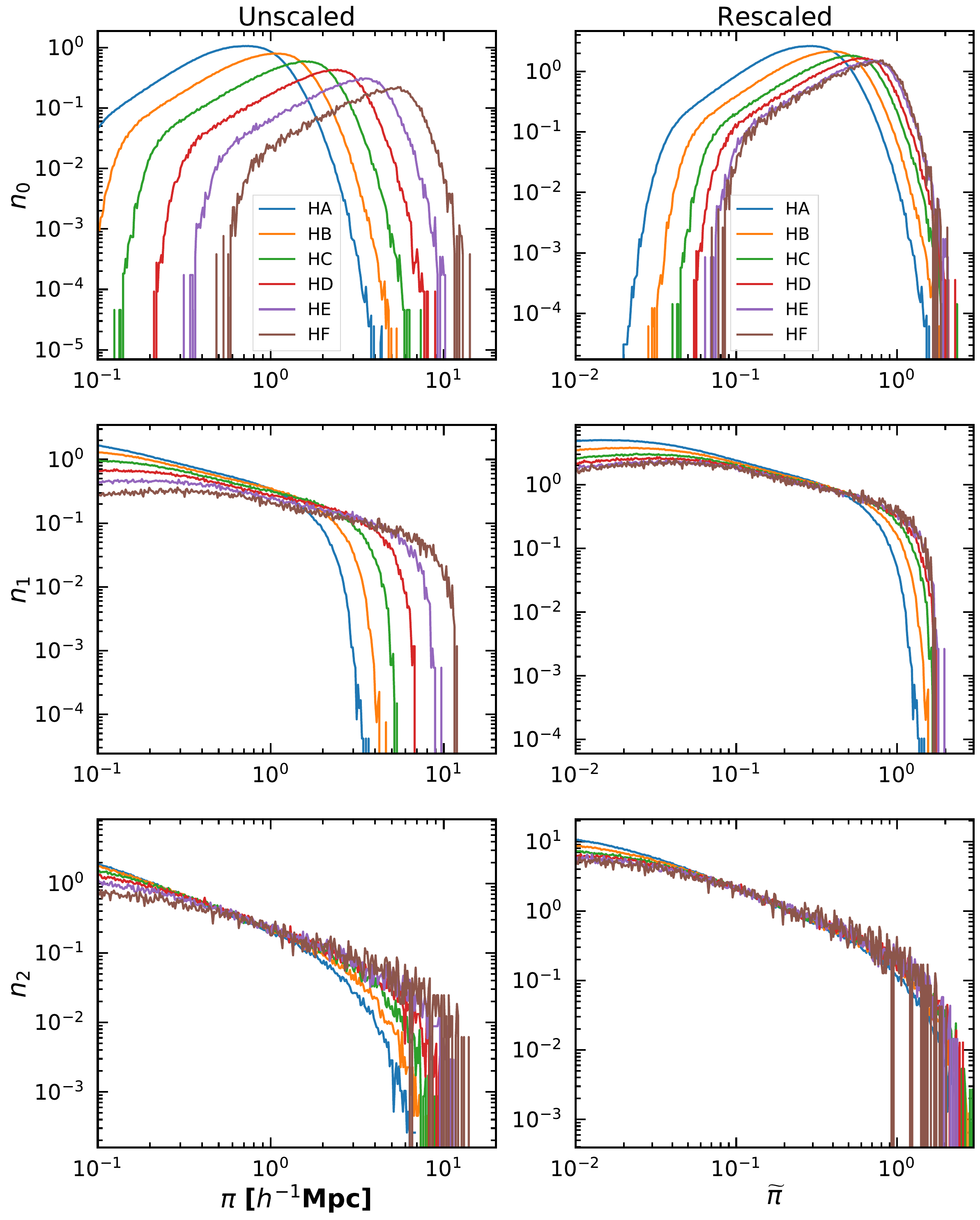}
\caption{\textbf{Persistence lifetime curves for all halo populations in the P-Millennium.} The curves show the fraction of persistence points $n_d$ at a specific persistence value (as defined in equation~\ref{eq:persistence}). From top to bottom, we show the curves corresponding to disconnected haloes, filaments and voids. The left column corresponds to the unscaled persistence curves, where $\alpha$ is in physical units of $h^{-1}$Mpc, whereas the right column corresponds to the re-scaled lifetime curves in units of $r_0$
(see Table~\ref{tab:Clustering Table} for exact values of $r_0$). Note that even when we take out the two-point correlation function from the persistence date (by considering the evolution of the alpha shape in terms of the scale-independent parameter $\widetilde{\alpha}= \frac{\alpha}{r_0}$), the haloes have a slightly different persistence. More remarkably, we see that the filaments (middle row) have a much sharper cut-off than the voids (bottom row). This is reflected already in the intensity persistence diagrams (see Figs.~\ref{fig:6_intensity_persistence}), the voids have a more gradual decrease in high persistence features, whereas the filaments have a much more clear break in the most persistent features.}\label{fig:7_lifetime_curves}
\end{figure*}

Finally, another different perspective for analysing persistence diagrams can be achieved with lifetime curves. These summary curves focus on a different aspect of persistence than the Betti curves, namely the persistence of a given feature. Each persistence value corresponds to a specific difference in birth- and death value, thus this style of depiction allows to analyse the multi-scale nature of the halo population with a focus on the stability of features and their distribution. In Fig.~\ref{fig:7_lifetime_curves} we show the relative prominence of features with a specific persistence by normalising the number of features at a specific persistence. The left column uses persistence based on the distance in physical units of $h^{-1}$Mpc, whereas in the right column the unscaled persistence (in units of $r_0$, exact value in Table~\ref{tab:Clustering Table}) is used, correcting for clustering according to the two-point correlation function. The rows show the behaviour of a different dimension of features, from top to bottom these are clusters of haloes, filaments, and voids. In all panels we focus on the higher ranges of persistence (i.e., longer-lived features above 0.1 $h^{-1}$Mpc/0.01$\cdot r_0$).

The unscaled persistence curves of the left column immediately highlight that features with higher persistence in all dimensions are more directly associated with heavier haloes, as indicated by the sharp cut-off increasing to a higher persistence threshold with higher halo mass. This both shows the more clustered nature of heavier haloes, but also their generally lower total number. The difference in cut-off is also notably stronger for zero- and one dimensional features, and would thus be visible in the halo clusters and the filamentary structure outlined by the different halo classes. The behaviour of the persistence curves is perceptibly (and unsurprisingly) more similar once we take clustering into account. The sharp cut-offs now occur at very similar values slightly above a persistence value of 1, and in particular the two-dimensional persistence curves start to exhibit almost self-similar behaviour when re-scaled. However, some deviations between the mass classes of haloes still remain, and the general trend of the unscaled curves is unchanged, with lower mass haloes having fewer prominent high persistence features.

To conclude, methods of persistent homology tell us that high persistence features trace the most prominent components, usually spanned by the most massive haloes. Analysing the persistence of the cosmic web is possible on three layers of detail, starting from Betti curves, the two-dimensional intensity persistence diagrams, and ending with persistence curves. Where before we used these methods to directly investigate the structure found in different mass classes of dark matter haloes, we now focus on the systematic trends that were observable. We will quantify these trends and highlight the associated topological bias.

\begin{figure*}
\centering
\includegraphics[width=1\textwidth]{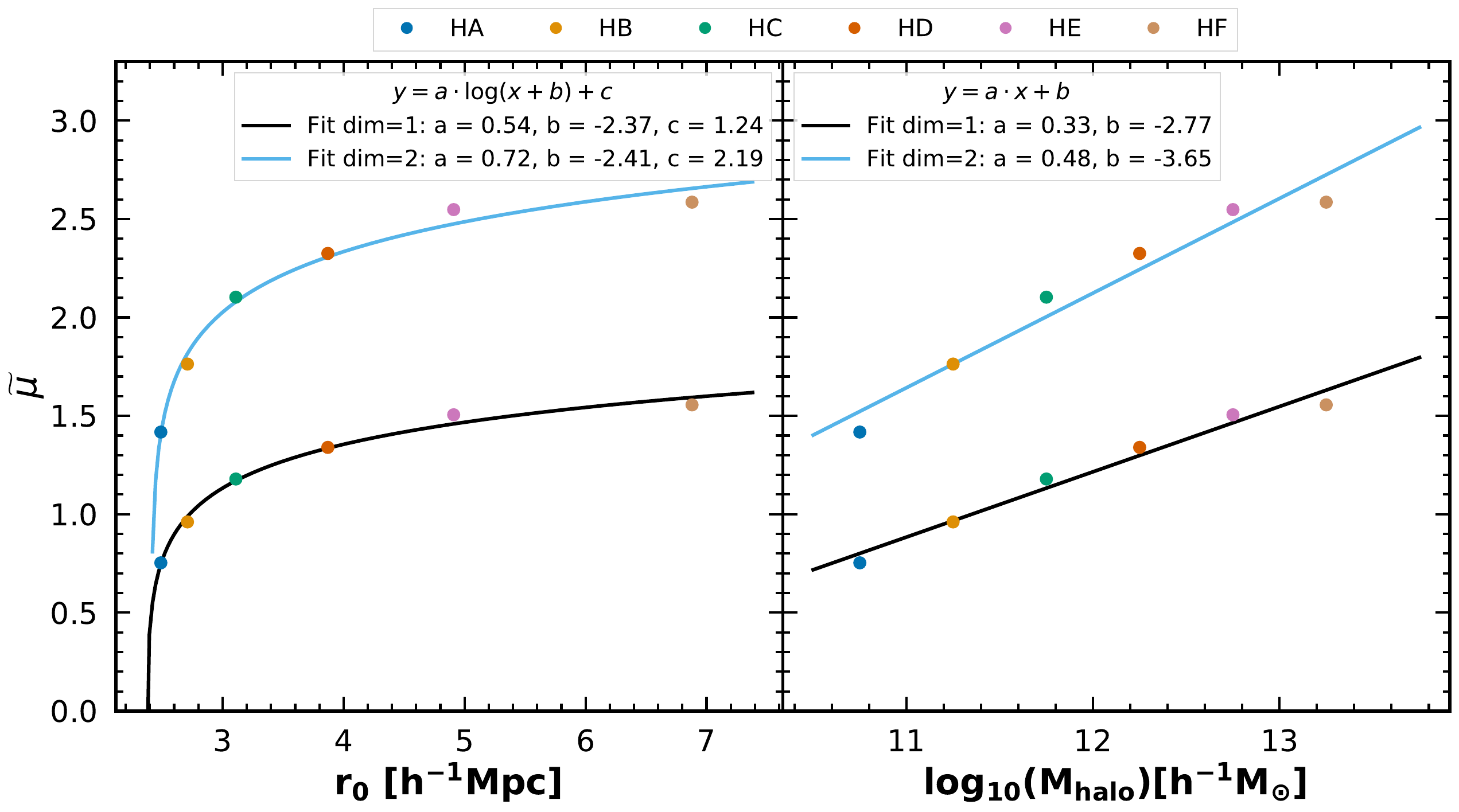}
\caption{\textbf{Betti curve scaling: dependence Betti mean on P-Millennium halo population.} We show the mean $\mu$ of the Betti curves (in each respective dimension) against the clustering length $r_0$ (left panel) and the halo bin mass (right panel). We model the behaviour of the Betti curves' modes depending on the reference frame by a natural logarithmic or a linear model (indicated in the legend).}\label{fig:8_bettimeanr0}
\end{figure*}

\subsection{Global Topology: Betti curves}
First, we quantify the location of the maxima of the Betti curves (see Fig.~\ref{fig:5_BettiCurves}) -- and thus any displacement and corresponding change of the characteristic length scales it encodes -- through its
mean, which is a good tracer of the location of the maximum of the curve (i.e., the mode). To obtain the mean we fit a skew normal distribution~\citep{ohagan1976bayes, azzalini1985class} to the Betti curves. We use this distribution to represent Betti curves also in earlier works~\citep[see, e.g.,][]{pranav2015thesis}, and in particular to parametrize the evolution of the cosmic web topology in a $\Lambda$CDM universe~\citep{wilding2021}. It is defined as
\begin{equation}\label{eq:skewnormal}
    f(x | \xi, \omega, \alpha, c) = \frac{2c}{\omega}\phi\left( \frac{x - \xi}{\omega}\right)\Phi\left(\alpha\frac{x - \xi}{\omega} \right)\,,
\end{equation}
where the function $\phi(x)$ is the unit variance zero mean normal probability density distribution defined on the real line, and $\Phi(x)$ is its cumulative distribution. The variable $x$ corresponds to the filtration value (e.g., $\alpha$), $\xi$ is a location parameter, $\omega$ a shape parameter, and $c$ is a normalisation constant. The mean $\mu$ can be obtained from the above fitting parameters through the relation
\begin{equation}
\mu = \xi + \omega\frac{2}{\pi}\delta  \,,\qquad \delta = \frac{\alpha}{\sqrt{1 + \alpha^2}}\,.
\end{equation}
In Fig.~\ref{fig:8_bettimeanr0} we show the influence of the halo mass influences on the mean of the scaled Betti curve, thus corrected for clustering. To investigate the relation to the halo mass itself, we show the change of the mean both depending on the clustering length $r_0$ of a halo mass class, as well as the relation to the halo mass itself. Depending on this reference frame we model the behaviour using either a logarithmic or a linear function. The parameters of the fits are given in the legend of Fig.~\ref{fig:8_bettimeanr0}. The parameter $\mu$ including uncertainties is given in Table~\ref{tab:linfit_mu}.

In both cases we observe clear trends between the scaled mean and the mass of a specific halo bin.
Halos in particular mass ranges form structures and features with an underlying topology that is not simply a differently-scaled version of haloes with a lower or higher mass.
Instead we see a distinctly different (topological) character, that is, a difference in prominence of one- and two-dimensional topological features depending on the mass range. This effect is apparent through the one- and two-dimensional trends differing in their slope, which points to a stronger correlation of halo mass and the two-dimensional topological features. These structures -- haloes in walls that surround voids -- experience a stronger change when looking at different halo masses than the one-dimensional filamentary structures.

\subsection{Multiscale Topology: Persistence diagrams}
%%% -------------------------------------------------------------

\begin{figure*}
\centering
\includegraphics[width=1\textwidth]{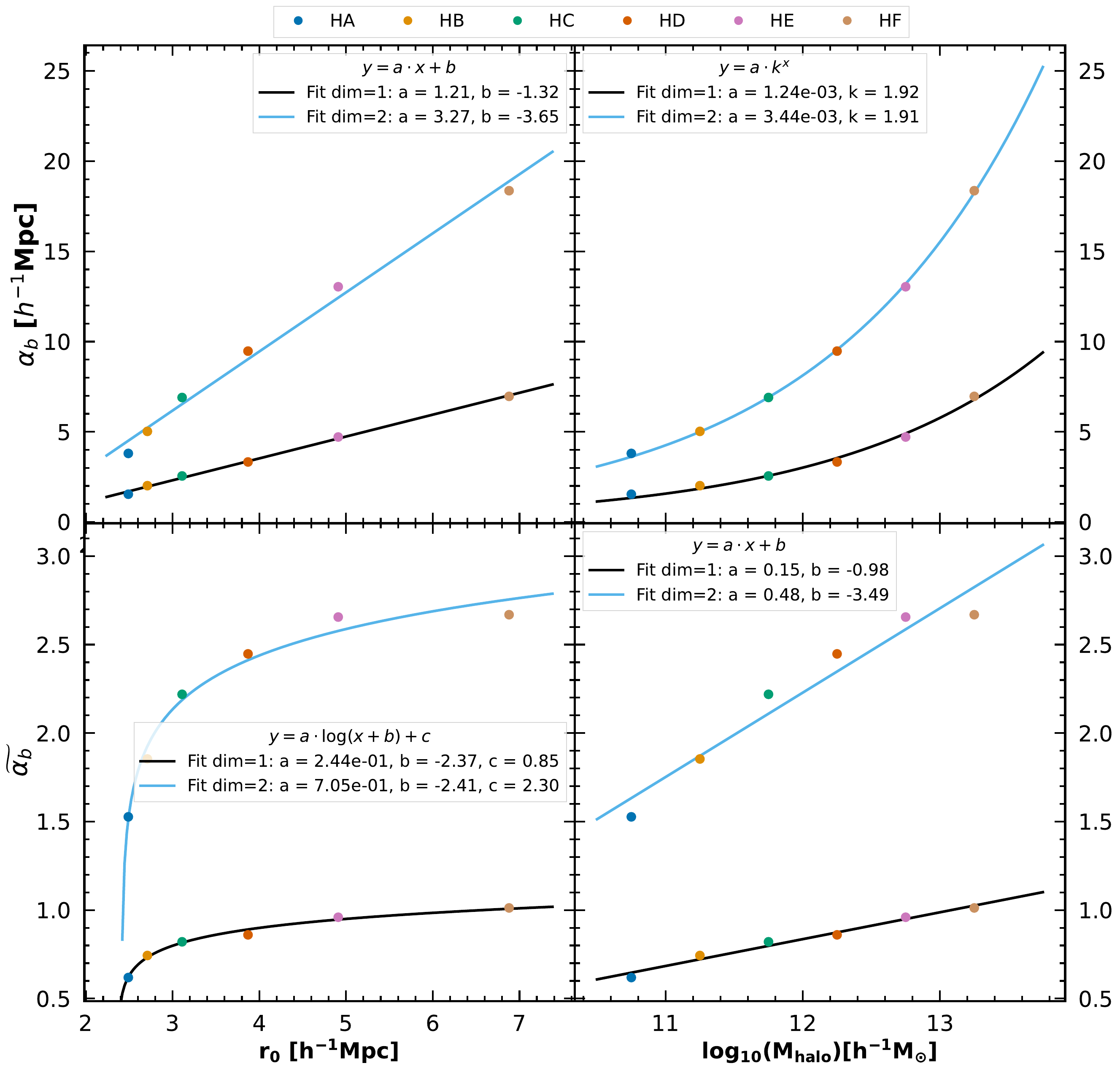}
\caption{\textbf{Apex of the intensity persistence diagrams in the P-Millennium.} Here, the apex of the triangle-like shape in the intensity persistence diagrams (as shown in Fig.~\ref{fig:6_intensity_persistence}), characterised through its birth value, is plotted as a function of the clustering length of each halo populations (left column), as well as the bin mass (right column). We show the unscaled apex in the top row and the scaled apex in the bottom row. The birth value $\alpha_b$ of the apex represents a turning point in the topology of different halo populations, both for filaments (dimension one) and voids (dimension two).
Depending on the reference frame, we use linear, logarithmic, or power law models (indicated in the legend) as bases to determine fitting parameters.}\label{fig:9_apexbirth}
\end{figure*}

We identify a second topological tracer through the persistence diagrams by looking at the location of the high-persistence apex. This region characterises long-lived, stable and prominent topological features, and is less prone to influence from short-lived and noisy features. The apex represents a highly relevant transition region in the topology of the halo population. For a sharp and narrow apex, e.g. visible in the two dimensional persistence diagrams in Fig.~\ref{fig:6_intensity_persistence}, this means the persistent structures are created at similar length scales, i.e. within a small range of values of $\alpha_b$. The apex thus represents a meaningful point for the change in the topology of these haloes, as features formed at both lower and higher values of $\alpha_b$ have a shorter persistence. Tracing its location (across different dimensions and halo populations) therefore allows to quantify these persisting structures themselves.

For each halo population, we compute $\alpha_b$ by averaging over the 25 pairs with the highest persistence $\pi = \alpha_d - \alpha_b$. In Fig.~\ref{fig:9_apexbirth}, we show the dependence of the apexes' birth value (as a quantifier for the location) on the mass of the halo bin. The top row uses the unscaled (not corrected for clustering) location. It indicates the connection length $\alpha$ in (comoving Megaparsec) at which the most persistent topological features of halo-structure form. The exact behaviour differs for the dimension in question. As expected, the formation of a filamentary halo structure (one dimensional, in blue) consistently occurs at a lower connection length than the formation of a void-like halo-structure (two dimensional, in grey), as the density in filaments is higher than in voids, and the distances between haloes shorter. When considering the location of the unscaled apex, the trends are well represented either by a linear fit when dependant on the clustering length (Fig.~\ref{fig:9_apexbirth}, top left), or by an exponential fit when dependant on halo mass (top right). The observed trends are similar to the ones for the mean of the Betti curves: a higher halo mass (and thus a higher clustering length) corresponds to a higher birth scale of the apex. The effect is again stronger for haloes weaving the void-like structure.

When we normalise for clustering, the influence of the halo mass on the detected topology is slightly reduced, but clear trends are still discernible (Fig.~\ref{fig:9_apexbirth}, bottom row). Where we previously used a linear and an exponential model, we now use a logarithmic model for the relation to the clustering length (bottom left), and a linear model for the relation to the halo mass (bottom right). In this representation, the similarity to the mean of the Betti curves as a topological tracer is particularly clear.
Once normalising second-order clustering, we are able to identify structure beyond the two-point correlation function and quantify it using topological tracers.

\begin{table}
    \centering
    \caption{Topological bias: linear relation between the logarithmic halo mass and mean of the rescaled Betti curve ($\widetilde{\mu}$, first and second row, see Fig.~\ref{fig:8_bettimeanr0}), and the apex (rescaled birth value $\widetilde{\alpha_b}$, third and fourth row, see Fig.~\ref{fig:9_apexbirth}).}
    \begin{tabular}{clcc}
                           & &   a       &       b \\ \hline
\multirow{2}{*}{Betti curve ($\widetilde{\mu}$) }  &  Filamentary structure   &   
0.33  $\pm$  0.03     & -2.8  $\pm$  0.3 \\
                    &  Void structure          &   
0.48  $\pm$  0.05       & -3.6  $\pm$  0.6 \\ \hline

\multirow{2}{*}{Apex ($\widetilde{\alpha_b}$)}  &  Filamentary structure   &   
0.15  $\pm$  0.01     & -1.0  $\pm$  0.1 \\
                    &  Void structure          &   
0.48  $\pm$  0.06       & -3.5  $\pm$  0.7
    \end{tabular}
    \label{tab:linfit_mu}
\end{table}

% Mean dim 1:
% a = 0.33185981214374416  $\pm$  0.027667142885614237
% b = -2.7666907500772675  $\pm$  0.33284523395270577
% Mean dim 2:
% a = 0.4810327405600282  $\pm$  0.0539381921748077
% b = -3.648639619423535  $\pm$  0.6488949821393086
% Mean dim 2 (restrict):
% a = 0.5646020130584359  $\pm$  0.036831661555763014
% b = -4.602722147090627  $\pm$  0.4335549629761386

% Apex dim 1
% a = 0.15165470472414044  $\pm$  0.011178347151708058
% b = -0.983988339828591  $\pm$  0.1344793573199821
% Apex dim 2
% a = 0.4767926323166843  $\pm$  0.0597771664897387
% b = -3.492817906621493  $\pm$  0.719139844091338
% Apex dim 2 restricted
% a = 0.5702524648046894  $\pm$  0.0392942141173468
% b = -4.559817660271652  $\pm$  0.4625423049503263

Focusing on halo mass as a discriminating attribute, we see that for most halo masses, the filamentary structure is formed through connections with a length of below one clustering length (between 0.5 and 1 $r_0$ for HA - HE, and slightly above 1 for HF). We note that the behaviour of the filamentary structure when expressed through the location of the apex is different compared to the mean of the Betti curves, and likely more accurate due to less influence of short-lived features. The void structure forms over a wider range of connection lengths, similar to what is apparent from the Betti curves (between 1.5 and 3 $r_0$).

\subsection{Topological Bias: quantitative relationships}
%%% -------------------------------------------------------------
To summarise, both topological tracers we introduce above quantify a topological bias we observe in the correlation between the mass of dark matter haloes and their geometrical environment. This correlation can be identified through the influence that a change in the mass of the haloes used to sample the cosmic web has on the persistent, multi-scale structure. The persistent structure is represented through persistent diagrams, in which we also pick up any change to topological buildup (e.g. in the connectivity) caused by a different sampling of the cosmic web. For a structure that features only second-order clustering, i.e., that is completely characterised by its two-point correlation function, but where clustering length still changes with the mass class, rescaling with the clustering length will remove any mass dependence of clustering. In this case, we would also not expect to pick up any change of the topological characteristics, i.e., we would expect a constant relation for the all curves shown in Figs.~\ref{fig:8_bettimeanr0} and \ref{fig:9_apexbirth}.

Since we do pick up clear trends, the clustering that is intrinsic to the halo distribution is different not only between different mass classes, but also exhibits distinct higher orders of clustering. The appearance of these higher orders of clustering is of course not surprising -- quite the contrary. However, the use of topological methods offers a novel approach at quantifying the complex clustering within the large-scale structure, different from using, for example, the three- or four-point correlation function. Instead of adding more points which can correlate, we include the rich topological information to determine the relation between the nature of clustering and an intrinsic structural dimension of the features under investigation. This allows us to observe that the mass has not the same influence on one-dimensional topological features (filamentary loops) as on two-dimensional features (voids). Instead we observe that the change in topology is stronger (the slope in Figs.~\ref{fig:8_bettimeanr0} and \ref{fig:9_apexbirth} is higher) for the latter -- the mass of a dark matter halo has a larger influence on its embedding in the void structure. Haloes of low to medium mass are more evenly distributed in the filamentary structure irrespective of the prominence (persistence) of features, whereas heavy haloes are bound stronger to prominent features in the void structure.

%########################################################
%                         SUMMARY
%########################################################
\section{Summary and conclusions} \label{sec: Conclusion}
The present study investigates the multi-scale topology of the spatial web-like pattern of the cosmic web, in an $\Lambda$CDM cosmological context, as traced by the population of dark matter haloes. This involves the question in how far the filaments and voids seen in the halo and galaxy distribution is representative for those present in the underlying dark matter web-like network.

Betti numbers quantify the number and prominence of voids, filaments and tunnels, and supercluster complexes in the matter distribution, and reflect the connectivity of these topological features. The multi-scale structure and connectivity of the various topological features is studied by means of persistence, in which one follows the scale at which individual topological features emerge, and the scale at which such features disappear as they merge with other features. The systematic inventory of these events, in the form of the different dimensional persistence diagrams, offers a direct view on the multi-scale nature of the cosmic mass distribution, its various morphological constituents and the nature of the hierarchical buildup of structure in the Universe.

In a range of previous publications we developed the use of homology and persistence to the analysis of the primordial matter distribution and the cosmic web~\citep{weygaert2011alpha,pranav2017topology,feldbrugge2019,wilding2021}. In \citet{wilding2021} we followed the topological evolution of the cosmic web in $\Lambda$CDM simulations, and established the close link between topological structure and key transitions in the cosmic structure formation process. The study provides a rich representation of the intricate and complex nature of the multi-scale web-like network in which dark matter has organized itself as a result of the gravitational evolution of the primordial density and velocity field.

Here we investigate the systematic dependence of the topology of the corresponding spatial halo distribution on the intrinsic properties of the haloes, specifically that of the mass of the haloes. The intention is to explore in how far the discrete sparse sample of haloes, and implicitly that of observed galaxies in galaxy surveys, reflects the intricate web-like structural pattern of the underlying dark matter distribution. In other words, in how far the topological structure found in the halo or galaxy population represents that of the dark matter distribution, and in how far it is significantly different. The presented topological analysis addresses the homology of the dark halo distribution in terms of the Betti numbers as a function of spatial scale, and of the corresponding persistence characteristics. 

\subsection{Halo sample}
%%% -------------------------------------------------------------
The population of dark haloes in this study stems from the Planck-Millennium~\citep{Baugh2019} (also see \citet{Punya2018}), a state-of-the-art dark matter particle simulation in a periodic box of side length $542.16 \ h^{-1}$Mpc. In this study we consider haloes with masses in excess of $3.2 \times 10^{10}\ h^{-1}M_{\odot}$, involving a sample of $1.13\times10^7$ haloes.

To assess the halo mass dependence of the topology and persistence of the P-Millennium halo sample, we extract six mass-defined halo subsamples. The subsample with the lowest mass haloes is that of haloes with mass between $10^{10.5}-10^{11}\ h^{-1}M_{\odot}$, the one with the highest mass haloes comprises haloes in the mass range $10^{13.0}-10^{13.5}\ h^{-1}M_{\odot}$. For each of the subsamples, we carry out a full computational Betti number and persistent homology analysis.

\subsection{Alpha shapes}
A key aspect of our homology analysis is that we study the discrete spatial halo distribution \emph{directly} as a function of \emph{spatial scale}, translating into a computational scheme based on \emph{alpha shapes}~\citep{Edelsbrunner1983,edelsbrunner1994}. These \emph{simplicial complexes} -- consisting of cells, faces, edges and vertices -- are subsets of the Delaunay tessellation defined by the scale $\alpha$.

The principal purpose of our study is to assess the structural scale dependence of (super)clusters, filaments and their associated tunnels, and voids. While most previous topology studies involved the dark matter \emph{density field}, to avoid the intermediate step of inferring the corresponding density field we work directly via the \emph{distance field} defined by the halo samples. For a discrete point sample it is in fact most efficient to base the computational analysis on a \emph{simplicial complex} that is topologically equivalent to the corresponding continuous distance field~\citep[see, e.g.,][]{edelsbrunner2010computational,boisonnat2018,carlsson2021}. To this end we invoke the Delaunay tessellation generated by the spatial halo distribution~\citep{delone1934sphere,okabe2000,weygaert1994fragmenting} to offer a discrete representation of the distance field. 

Probing the full multi-scale topology of the halo distribution involves the study of all individual changes in topology over an entire range of spatial scales. It naturally leads to the concept of \emph{distance field filtrations}, and within its simplicial representation to that of \emph{alpha shapes}. The filtration of a field is a subset of the field (manifold) that obeys a specific condition: usually it is the superlevel or sublevel set defined by a certain threshold. Applied to the Delaunay tessellation, the $\alpha$ \emph{shape} is a simplicial filtration defined by a scale $\alpha$, whose simplicial elements are the Delaunay simplices that represent connections up to the distance threshold $\alpha$. Changes in the topology occur when larger Delaunay simplices get incorporated as the distance threshold increases.

\subsection{Topology of the halo-sampled cosmic web}
%%% -------------------------------------------------------------
The topological analysis of the six mass-based halo samples reveals a clear mass dependence. For all three Betti number curves $\beta_0(\alpha)$, $\beta_1(\alpha)$, and $\beta_2(\alpha)$ we find a stark and systematic difference between the different halo samples. As the halo mass increases, we see the Betti curves become broader and extend out to larger scales $\alpha$. We also see that the amplitude of the $\beta_1$ and $\beta_2$ curves decrease systematically as the halo mass increases.

The systematic shift in characteristic topological scale is a direct consequence of the spatial structures delineated by the haloes in the corresponding mass range. The higher mass haloes delineate voids and filaments of a larger scale than those traced by the lower mass haloes. The characteristic scale of filaments shifts from around $2$ for the lowest halo mass range to around $10 \ h^{-1}$Mpc for the highest mass haloes. Voids delineated by the halo distribution have a spatial scale that is roughly twice as large, ranging from $4$ for the low halo mass range to $20 \ h^{-1}$Mpc for the high mass haloes. The increasing size of the filaments and voids goes along with a lowering of their total number. In other words, high mass haloes trace the fewer and larger specimen of these structural features. 

This is, as expected, also mirrored in similar scale shifts in the persistence diagrams. Overall, the topological features traced by high mass haloes are born and disappear at larger scales. Particularly outstanding are the persisting features in the diagrams. The apex in the $\beta_0$ diagrams, the sharp maximum that marks the scale at which the most dominant voids emerge, also sees a similar shift to larger scales as the halo mass increases. The immediate implication is that most voids in higher halo mass samples emerge at systematically larger scales.

\subsection{Sampling and clustering}
The fact that the higher mass haloes only trace larger filaments and voids, while missing out on the
smaller and more tenuous ones, is not unexpected.

One factor is that of the lower abundance of higher mass haloes. As a result of the larger mean halo distance of the high mass sample, they are not able to trace small filaments and voids. In other words, the high mass halo samples offer a sparser and more diluted view of the overall structure. A physically more significant effect is that of the clustering of the various halo samples. It is known that higher mass haloes are intrinsically more strongly clustered~\citep{Kaiser1984}. The stronger clustering translates into a more clumpy distribution. By default, this will entail a pattern marked by large-scale features, such as larger underdense regions and hence larger voids. 

To first approximation, the strength and scale of clustering is encapsulated in the second-order -- two-point -- correlation function. However, complex structural patterns such as those seen in the cosmic web and which here we seek to describe topologically entail higher order contributions. Given our interest in significant differences in the intrinsic topology of the web-like patterns traced by the different halo samples, we therefore need to compensate for the clumpiness and scale dependence implied by the two-point correlation function of the halo samples.

The two-point correlation functions of the halo samples are simple power-laws, whose sole difference is that of the specified amplitude in terms of the clustering length $r_0$. Hence, to remove the effect of the differences in two-point correlation functions between the different halo samples we renormalise the scales $\alpha$ in each halo sample by the corresponding clustering length, ${\tilde \alpha}=\alpha/r_0$.

A major asset of the rescaling is that it enables the comparison on an equal footing of the spatial patterns outlined by the halo populations. Differences in rescaled spatial structure are manifestations of the presence of higher order correlations: the two-point correlation function does not entail any phase correlations and is therefore insensitive to the presence of complex spatial patterns in the distribution of haloes~\citep{coles2000,coles2009}. It is the rich topological language of (persistent) homology that offers a visually and physically accessible characterisation of these higher order influences.

\subsection{Topological bias: Definition}
%%% -------------------------------------------------------------
The major finding of our analysis is that stark and systematic mass dependent differences in topological structure remain between the renormalised halo samples. If anything, they are even more outstanding, following strict linear relations between topological quantities and (logarithmic) halo mass. The significant implication is that the web-like patterns and structures, such as filaments and voids, probed by haloes of different mass differ fundamentally.

For the rescaled halo samples, we observe that the Betti number curves $\beta_0(\tilde \alpha),~\beta_1(\tilde \alpha),$ and $\beta_2(\tilde \alpha)$ (see Fig.~\ref{fig:5_BettiCurves}) systematically shift towards higher (renormalised) scales as a function of halo mass:
\begin{itemize}
\item  Light haloes, less than $10^{12} M_\odot$, arrange themselves along filamentary structure on scales \emph{less than} the clustering length.
\item Higher mass haloes trace filaments on scales larger than $r_0$. This also holds for voids, for haloes over the entire mass range.
\item However, the rescaled Betti curves hardly broaden, while the amplitude of the $\beta_1(\tilde \alpha)$ curves for filaments and tunnels only decreases slightly, and remains roughly similar for the void $\beta_2(\tilde \alpha)$ curves.
\item Hence, in terms or renormalised scale: with increasing halo mass we hardly see a change in the number of voids, along with a moderate decrease in the number of filaments.
\end{itemize}

\noindent The strong nature of the systematic topological differences between filaments and voids traced by low mass haloes and those traced by high mass haloes is underlined by unequivocal linear quantitative trends (see Figs.~\ref{fig:8_bettimeanr0} and \ref{fig:9_apexbirth}). We observe a linear relation depending on the logarithmic halo mass:
\begin{itemize}
\item  The mean (renormalised) spatial scale of the Betti curves, marking the average scale of the corresponding topological features.
\item  The (renormalised) spatial scale at the apex of the persistence diagrams for $d=1$ (filaments) and $d=2$ (voids). The apex of the persistence diagram marks the emergence (birth) of topological features, and encodes the hierarchical evolution that produced these.
\item The proportionality factor $a$ of the linear relations for one- and two-dimensional topological features differs: the relation is steeper for the $d=2$  situation of voids as compared to that for the $d=1$ situation for filaments and tunnels. 
\end{itemize}
These observations reveal the fundamental nature of the changing topological character of the spatial halo distributions. The immediate implication is that filaments and voids traced by haloes of different mass are fundamentally different, they differ in their topological character. Moreover, the dependence on halo mass in tracing the void population and their structure appears to be stronger than that for filaments.

The fact that the higher-order structure of the cosmic web depends so strongly on the tracing halo population, and that this finds its expression in strong linear relationships of topological quantities with the (logarithmic) halo mass, may be indicated as \emph{topological bias}. It means that haloes of different mass trace environments with different topological signature. It relates to a novel kind of halo bias, not related to second-order clustering nor directly to the underlying matter density but to the topological features in which they are located. Moreover, it affects more strongly the voids than the filaments in the cosmic web.

\subsection{Topological bias: Implications}
%%% -------------------------------------------------------------
\emph{Topological bias} specifies that filaments and voids fundamentally differ when seen in the spatial distribution of different populations of haloes and galaxies. It is a manifestation of fundamental differences between the landscapes of filaments, tunnels, walls, and voids that constitute the structural elements of the cosmic web.

Overall, we may understand \emph{topological bias} as the origin of the fundamental differences in structure between the filament and void population traced by high mass versus low mass haloes. Low mass haloes also populate the more tenuous structures in the cosmic mass distribution. Reflecting their hierarchical buildup, the filamentary network of the cosmic web branches into smaller tendrils~\citep[see e.g.][]{cautun2014}. The substructure of voids -- the \emph{void hierarchy} -- is the result of an even more complex hierarchical history~\citep[see e.g.][]{dubinski1993,shethwey2004,aragon2013}. While the low mass haloes manage to outline the finer features of the hierarchically evolved cosmic web, high mass haloes tend to exclusively concentrate along the largest structures in the cosmic web. A closer investigation of the origin and evolution of topological bias would shed light on the possible relation to other aspects of halo and galaxy bias, amongst which that of assembly bias~\citep{Gao2005,Wechsler2006,dalal2008,Mao2018}.

As a result, low mass haloes succeed in delineating the topological fine structure of the cosmic web. Evidently, the observational complication of finding low mass haloes and low luminosity galaxies will remain a major challenge in the ability to study in the observational reality the full richness of the topological structure of the cosmic web and topological bias. 

Topological bias will impact on a wide range of cosmologically interesting aspects of the cosmic web. It will influence the outcome of studies of the filament population~\citep{jones2010,tempel2014,libeskind2018}, and their use to infer intrinsic alignments between the spin of galaxies -- and other galaxy properties -- and the harbouring filaments~\citep[see e.g.][]{porciani2002,porciani2002b,aragon2007,schafer2009,jones2010,codis2012,tempel2013,alpaslan2014a,Jarrett2017,Kraljic2018,Punya2018,punya2021,welker2020}. It affects even more strongly the voids outlined by haloes, and should be taken into account when seeking to derive the size distribution of cosmic voids, and use
these as a measure of the underlying cosmology~\citep[see e.g.][]{shethwey2004,platen2008,weygaert2008clusters,alpaslan2014b,pisani2015,weygaert2016voids,verza2019,pisani2019cosmic}. 

In conclusion, the implications of the existence of topological bias will be particularly important for the analysis and interpretation of the spatial galaxy distribution in galaxy redshift surveys. This is true for existing surveys, such as 2dFGRS, SDSS, GAMA, 2MASS, and VIPERS~\citep{colless2003,tegmark2004cosmological,gama2009,  huchra20122mass,vipers2013} as well as for the upcoming major surveys that will map the cosmic web in unprecedented detail, such as DESI, Euclid, the Vera Rubin telescope and SKA related surveys.

\section*{Acknowledgements}
The authors would like to thank Carlos Frenk, for the access to the halo catalogue of the P-Millennium simulation and encouraging discussions. We are also very grateful to Carlton Baugh, Job Feldbrugge, Roi Kogul, Keimpe Nevenzeel, and Punya Ganeshaiah Veena for the many insightful discussions that were crucial towards the completion of this work. This project is conducted in part at the Centre for Data Science and Systems Complexity at the University of Groningen and is sponsored with a Marie Sk\l odowska-Curie COFUND grant, no. 754315.

\section*{Data availability}
The data underlying this article were provided by Carlos Frenk and will be shared on request to the corresponding author with his permission.

%%%%%%%%%%%%%%%%%%%%%%%%%%%%%%%%%%%%%%%%%%%%%%%%%%



\bibliographystyle{mnras}
\bibliography{references} % if your bibtex file is called example.bib

%%%%%%%%%%%%%%%%%%%%%%%%%%%%%%%%%%%%%%%%%%%%%%%%%%

%%%%%%%%%%%%%%%%% APPENDICES %%%%%%%%%%%%%%%%%%%%%

\appendix

\section[Simplicial complexes: Voronoi \& Delaunay tessellations]{\\ Simplicial complexes:\\ Voronoi \& Delaunay tessellations}
\label{app:simplicial}

The present study focuses on the multi-scale topology -- persistent homology -- of the halo distribution in the P-Millennium simulation. Given the discrete nature of the spatial halo distribution, we opt for a direct computation of the multi-scale topology of the corresponding point distribution. This is accomplished on the basis of the \emph{Delaunay tessellation}~\citep{delone1934sphere,okabe2000,edelsbrunner2010computational} generated by the point distribution. The Delaunay tessellation is a simplicial complex that is optimally sensitive to two key aspects of the cosmic web: the multi-scale nature of the structure probed by the point distribution and its anisotropic morphology/geometry, in particular that of the prominent filamentary network and the connected flattened walls~\citep[see][]{weygaert2008cosmic}.

For the assessment of the multi-scale topological structure of the cosmic web we need to invoke a well-defined scale-based filtration of the simplicial complex that represent the spatial structure of the point distribution. This leads directly to a key concept in computational topology and geometry, known as \emph{alpha shapes}~\citep{Edelsbrunner1983,edelsbrunner1994,edelsbrunner2010computational}. Uniquely defined for a particular point set $P$ by the scale parameter $\alpha$, alpha shapes correspond to a unique assembly of simplicial (geometric) structures that capture the shape and morphology of the point distribution.

This appendix provides definitions, details and illustrations of the related concepts from computational geometry and topology. Via illustrations of the 3D alpha shapes of the web-like distribution of haloes in cosmological simulation, we demonstrate the meticulous means by which alpha shapes are probing the topological nature of complex spatial patterns. 

\subsection{Simplicial complex}
A simplicial complex is an ordered geometric assembly of faces, edges, nodes, and cells that mark a discrete spatial map of the volume containing a point set~\citep{edelsbrunner2010computational,pranav2017topology}. The geometric components of the complex are simplices of different dimensions: a cell is a three-dimensional simplex, a face or wall a two-dimensional simplex, an edge a one-dimensional simplex and a node a zero-dimensional simplex. A good impression of a two-dimensional simplicial complex can be obtained from Fig.~\ref{fig:1_AlphaShape_halos}, showing a slice through the Delaunay tessellation defined by the spatial halo distribution in the P-Millennium simulation, and a range of its alpha shape filtrations (see Section~\ref{app:alphashape}). Zero-, one- and two-dimensional simplices are visible: they correspond to the vertices and edges of the particle distribution's tessellation.

Of key importance for our study is that when the simplicial complex is defined on the basis of a generating point set $P$, when the complex is volume-covering and reflecting the number density and geometry of $P$, the lengths of the edges in the complex represent a sampling of the corresponding distance field. We will exploit this with respect to the Voronoi and Delaunay tessellation of the point samples described in this study~\citep{Voronoi1908,delone1934sphere,okabe2000}.

\subsection{Delaunay \& Voronoi tessellations}
\label{app:vordeltess}
The first step of the computational procedure to analyse the halo topology is the simplicial representation of the halo distribution samples. The simplicial complex should represent the spatial characteristics of the point distribution. To optimally represent the multi-scale nature of the spatial point distributions, as well as its organisation into a network of anisotropic filamentary and flattened features, we invoke their Voronoi and Delaunay tessellations~\citep{Voronoi1908,delone1934sphere,Icke1987,weygaert1994fragmenting}. Of crucial importance is also that the Delaunay tessellation facilitates the definition of a transparent and uniquely defined simplicial \emph{filtration}, that of the \emph{alpha shapes}~\citep{Edelsbrunner1983,edelsbrunner1994,edelsbrunner2010computational} (see Section~\ref{app:alphashape}).

%%%%%%%%%%%%
%%% From p9 Section 2.5 of manuscript 
%%% Moved to Appendix B as new subsection Delaunay triangulation}
%%%%%%%%%%%%

The Delaunay and Voronoi tessellation are amongst the most well-known concepts in computational and stochastic geometry, uniquely defined for a spatial point distribution $P$ (see Appendix~\ref{app:vordeltess} for the detailed definition). Delaunay and Voronoi tessellations have a range of properties that make them ideally suited for studies of the cosmic web in galaxy surveys and computer simulations~\citep[see eg.][]{Icke1987,weygaert1991,weygaert1994fragmenting,okabe2000,weygaert2008cosmic,edelsbrunner2010computational}.

\subsubsection{Geometric definition}
The Voronoi tessellation~\citep{Voronoi1908} and Delaunay tessellation~\citep{delone1934sphere} are amongst the most well-known simplicial complexes~\citep[see][for an extensive overview]{okabe2000}. They are uniquely and fully defined by a given spatial point distribution (i.e., for a generic non-degenerate point set). The Voronoi tessellation $\mathcal{V}(P)$ of a point set $P$ divides up space into polyhedral cells (3D) or polygonal cells (2D). Each cell $\mathcal{V}_p$ allocates the part of space that is closer to a given point $p$ than
to any of the other points in the (finite) point sample $P\subseteq \mathbb{R}^3$. Formally, the Voronoi cell $\mathcal{V}_p$ of point $p$ is defined as:
\begin{equation}\label{eq:voronoicell}
    \mathcal{V}_p = \left\{x \in \mathbb{R}^3 \hspace{1mm} | \hspace{1mm} ||x-p|| \leq ||x-q||, \forall \hspace{1mm} q \in P \right\}.
\end{equation}
The \emph{Delaunay tessellation} is the dual of the \emph{Voronoi tessellation}, and the calculation of one of these tessellations immediately yields its dual. The Delaunay tessellation is the volume-filling triangulation consisting of (3D) tetrahedra whose four vertices are points of the sample $P$ whose circumscribing sphere does not contain any of the other points. In the two-dimensional situation this concerns triangles, defined by three points whose circumscribing circle does not contain any other points of the sample $P$. Each vertex in the Delaunay tessellation is a nucleus of a Voronoi cell. The \emph{edges} connect two sample points that share a common face in the Voronoi tessellation, constituting \emph{natural neighbours}. Also, the \emph{edges} of a Voronoi polyhedron pass through centroids of the triangular faces of a Delaunay tetrahedron, while the centroid of the Delaunay tetrahedron is one of the vertices of the polyhedral Voronoi cell.

There is a vast mathematical and computational literature on these volume-covering tessellations, and they know a wide range of applications in science and engineering. Many of the relevant aspects are treated in the volume by \citet{okabe2000}. 

\subsubsection{Tessellations and the cosmic web:\\ \ \ \ \ \ \ \ \ \ \ \ Sensitivity to multi-scale and anisotropic morphology}
A detailed assessment of the sensitivity of Voronoi and Delaunay tessellations to the nature of the generating point distribution~\citep{schaap2007,weygaert2008cosmic} revealed that they impressively follow the fundamental structural properties of the cosmic web. Voronoi cells are highly sensitive to the local number density of the point distribution~\citep[see][]{weygaert1991,weygaert1994fragmenting,schaap2000continuous,Neyrinck2005,weygaert2008cosmic}. The dual of the Voronoi tessellation, the Delaunay tessellation, shares these properties to an even stronger extent. 

The fully self-adaptive DTFE procedure exploits these properties~\citep{schaap2000continuous,schaap2007,weygaert2008cosmic,cautun2011}. It delivers an impressively accurate reconstruction of the spatial structure of the density field traced by the points sample. It uses the Voronoi cell volumes to obtain an estimate of the local density at each sample point, and subsequently uses the Delaunay cells to define a first-order estimate of the density field throughout the sample volume.The impressive ability of DTFE to recover the details of the \emph{multi-scale nature} and \emph{anisotropic geometry} of the cosmic web reflects the fact that the simplicial composition of the Delaunay tessellation itself is highly sensitive to these properties. A telling visual demonstration of this is shown in the Delaunay tessellation in and around a  filament in a cosmological simulation, illustrated in figure~17 in \citet{weygaert2008cosmic}.
A careful assessment of the properties of DTFE, the Delaunay Tessellation Field Estimator, revealed that it manages to trace accurately key characteristics of the structure of the cosmic web~\citep{weygaert2008cosmic}. To a considerable precision it follows the multi-scale nature of the mass distribution sampled by a point distribution. DTFE manages to do so to the extent that it even reproduces the fractal nature of a point distribution organized in a nested sequence of ever denser point clumps. Equally important is that DTFE manages to reproduce accurately the shape of the local point distribution, and hence manages to retain the planar geometry of wall-like distributions and the elongated nature of points organized along filaments. In addition, it yields
the corresponding volume-covering volume-weighted velocity field traced by the points~\citep{bernardeau1996,romano2007delaunay}.

\section{Alpha shapes}
\label{app:alphashape}
The self-adaptive nature of Delaunay tessellations with respect to local density and shape of a spatial point distribution assures that the key characteristics of the cosmic web are retained. This also concerns all information on the topology of the structure traced by the sample points. It naturally leads to the concept of the scale-based filtration of a Delaunay tessellation, defining simplicial complexes known as \emph{alpha shapes}. These form the basis for a well-defined framework for the analysis of the multi-scale geometric and topological nature of the spatial mass distribution sampled by the sample points~\citep{Edelsbrunner1983,edelsbrunner1994,edelsbrunner2010computational}.

Alpha shapes are a well-known concept in Computational Geometry and Computational Topology~\citep{dey1999, rote2006computational,zomorodian2005,edelsbrunner2010computational}. They were introduced by Edelsbrunner and collaborators~\citep{Edelsbrunner1983,edelsbrunner1994}. They involve a generalization of the convex hull of a point set and are concrete geometric objects that are uniquely defined for a particular point set on a scale parameterized by the scale parameter $\alpha$. They have found a wide array of scientific and engineering  applications, including pattern recognition, digital shape sampling and processing, and structural molecular biology~\citep[e.g.][]{liang1998a,liang1998b}. For example, in the latter they have been used in the characterisation of the topology and structure of macro-molecules. They were also central in our earliest studies of Betti number characteristics and persistent topology of the cosmic mass distribution and the cosmic web~\citep{eldering2005,weygaert2010alphashapetopology,weygaert2011alpha}.  

%%%%%%%%%%%%
%%% From p9 end Section 2.5 of manuscript 
%%% Moved to Appendix B as new subsection Delaunay triangulation}
%%% This is very similar to preceding text and couild be entirely eliminated
%%%%%%%%%%%% 

\subsection{Alpha shapes: Definition}
Alpha shapes form the core of a well-defined framework for the analysis of the multi-scale geometric and topological nature of the spatial mass distribution traced by the sample points~\citep{Edelsbrunner1983,edelsbrunner1994,edelsbrunner2010computational}. A visual impression of a two-dimensional alpha shape filtration -- for a distribution of haloes -- is provided by Fig.~\ref{fig:1_AlphaShape_halos}. Where for $\alpha=0$ (top left) we only observe the points without any connections, this quickly changes when we increase $\alpha$ to 3 or 6 $h^{-1}$Mpc (top centre and right) and see the emergence of a connected network. The illustration also elucidates the link between alpha shapes and the homology of a point distribution. For example, we see that tunnels are formed when, at a certain value of $\alpha$, an edge is added between two vertices that were already connected.

Rigorously defined in terms of the scale factors $\alpha$, alpha shapes are subsets of Delaunay tessellations that represent a scale-based filtration of the distance field. The Delaunay tessellation filtration by means of alpha shapes proceeds as follows. Centered around each generating sample point, i.e., around each vertex of the Delaunay tessellation, a sphere of radius $\alpha$ is drawn. Subsequently we identify the simplices of the corresponding Voronoi tessellation -- the dual of the Delaunay tessellation -- that are located within or intersect with the $alpha$ sphere union. The \emph{alpha complex} consists of all Delaunay simplices that record (are dual to) these Voronoi simplices. The \emph{alpha shape} for that specific value of $\alpha$ is the union of simplices in the alpha complex. For telling explanatory figures we refer to works by \citet{edelsbrunner2010computational} and \citet{weygaert2011alpha}.

\begin{figure*}
   \includegraphics[width=1\textwidth]{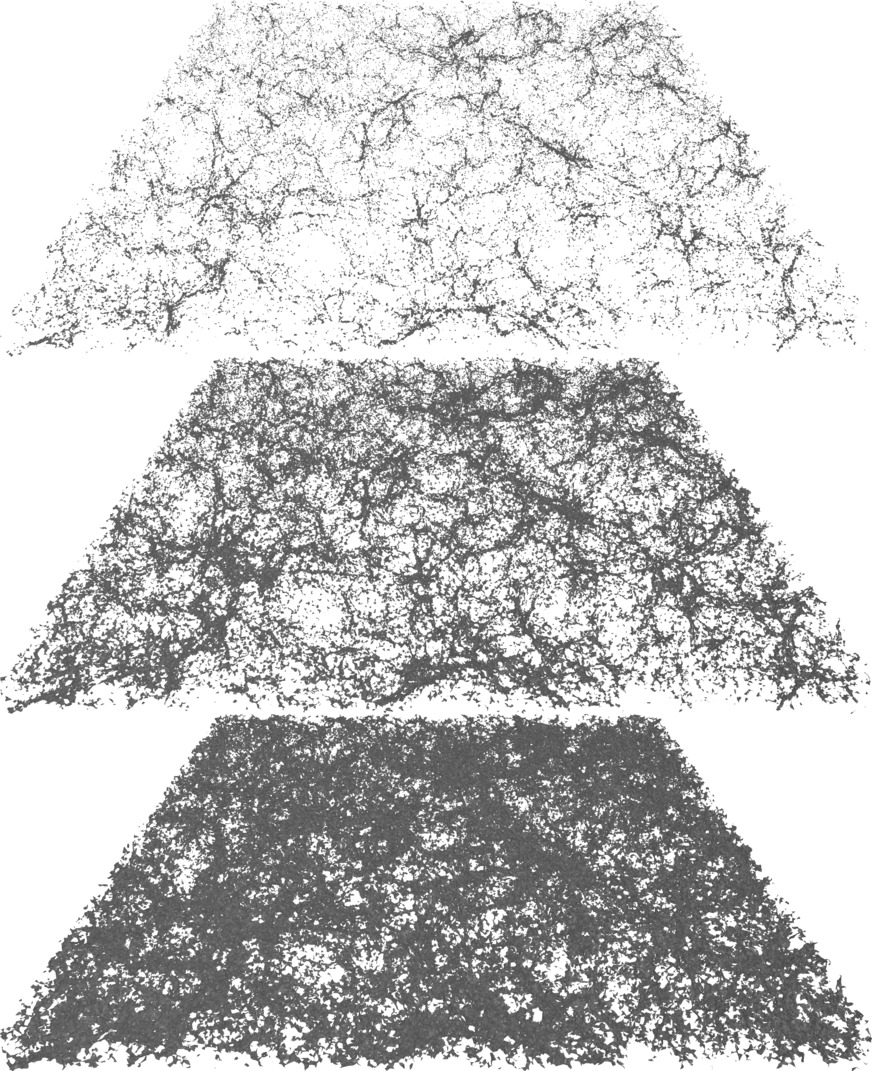}
\caption{\textbf{Three-dimensional alpha shapes.} From top to bottom we show the three-dimensional alpha shapes of a slice of the full halo distribution for $\alpha \approx 0.8$, $1.2$, and $1.6$ $h^{-1}$Mpc (corresponding to one third, one half, and two thirds of the average halo-halo distance).}\label{fig:B1_3d_alphashape}
\end{figure*}
% 0.6 = avg/4
% 0.8 = avg/3
% 1.2 = avg/2
% 1.6 = 2*avg/3
% 2.4 = avg

\begin{figure*}
   \includegraphics[width=1\textwidth]{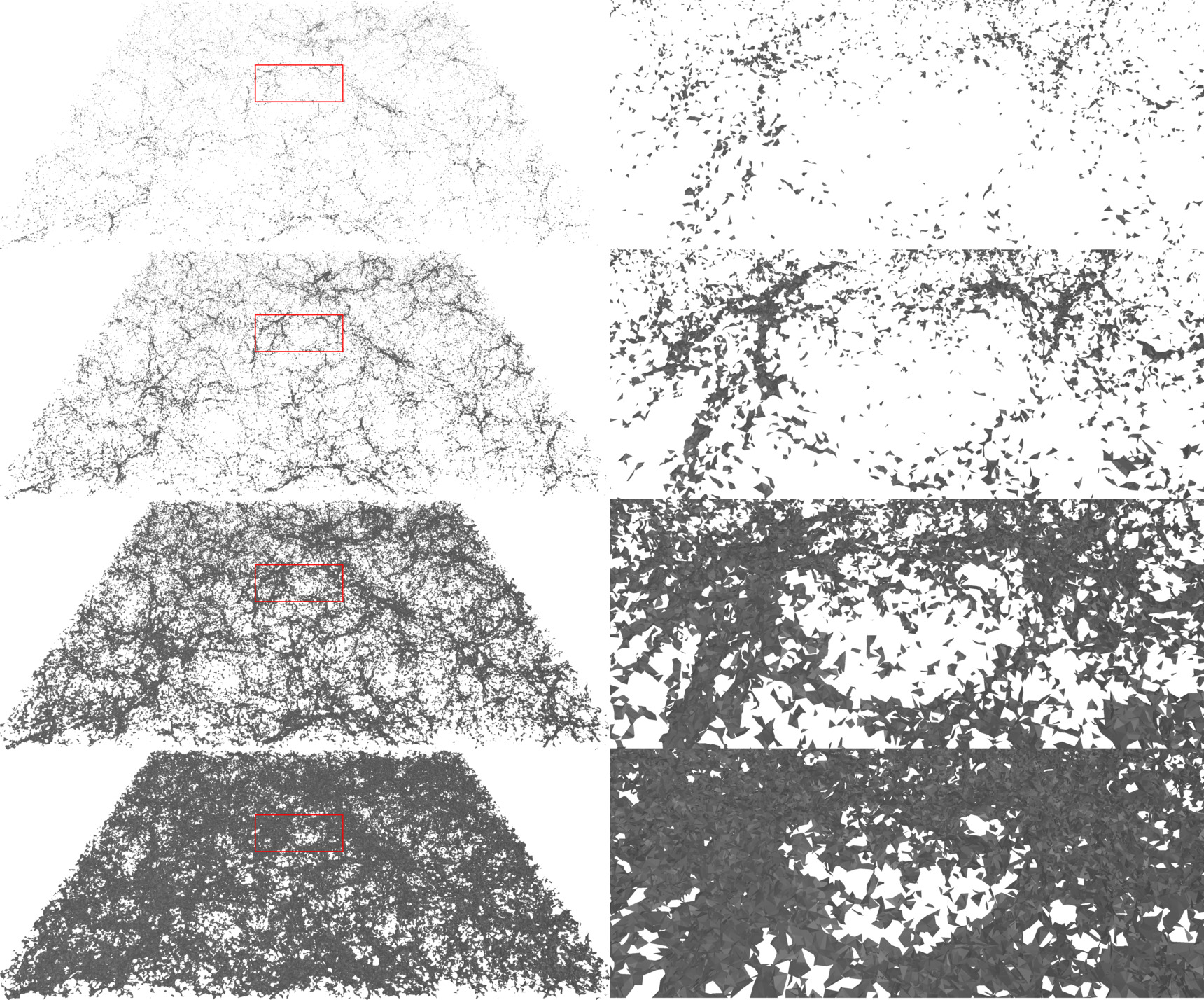}
\caption{\textbf{Three-dimensional alpha shapes.} From top to bottom we show the three-dimensional alpha shapes of a slice of the full halo distribution for $\alpha \approx 0.6$, $0.8$, $1.2$, and $1.6$ $h^{-1}$Mpc (with the additional top slice corresponding to a quarter of the average halo-halo distance). The right-hand column shows a zoom-in of roughly 100 $h^{-1}$Mpc width, which is marked by a red rectangle in the corresponding left-hand panel.}\label{fig:B2_3d_alphashape_zoom}
\end{figure*}

\subsection[Alpha shapes: Delaunay filtrations \& persistent topology]{Alpha shapes: \\ \ \ \ \ \ \ \ \ \ Delaunay filtrations \& persistent topology}
The ordered set of alpha shapes constitute a distance filtration of the Delaunay simplicial complex $\mathcal{D}$
\begin{equation}
 \emptyset = \mathcal{D}_0 \subseteq \mathcal{D}_1 \subseteq \mathcal{D}_2 \subseteq ... \subseteq \mathcal{D}_m = \mathcal{D}.
\end{equation}
The alpha shapes $\mathcal{D}_k$ are hierarchical subsets of $\mathcal{D}$ that geometrically encode the change in topology throughout the simulation box, and constitute the \emph{alpha complex}. Important is that they are homotopy equivalent to the distance field. It means that the topology of the alpha shape simplicial complex is topologically entirely equivalent to the continuous distance field~\citep{weygaert2011alpha}.

Instead of studying the topology at a single length scale, alpha shape topology probes the topology of the distance field along its set of filtrations along the varying length scale $\alpha$. The topological information in the simplicial alpha complex is extracted by assessing the connectivity of the various simplicial components. By following how the connections between points and simplices -- within the setting of the corresponding Voronoi and Delaunay tessellation -- depend on the scale parameter $\alpha$ we arrive at a natural description and rigorous definition of the multi-scale topology. As the filtration length scale $\alpha$ increases, the sample point distribution proceeds through a process in which we see a growing number of Delaunay simplices getting connected as they get embedded in the corresponding alpha complex. The increasing filtration length involves a continuously changing population of different separate simplicial complexes and the corresponding structural entities they represent.

Proceeding through the entire value range of the scale parameter $\alpha$, we may follow the creation and destruction of the individual topological features. As new structures emerge, existing structures merge and other structures get annihilated, the number of (super)clusters of haloes, filaments and voids will continuously change. To obtain insight into the connection between the growing simplicial (alpha shape) complex and the topology of the spatial structure, it is illustrative to consider what happens as more points get added while the simplicial complex grows~\citep[see][]{weygaert2011alpha}. Cycling through the simplices of the alpha complex, we encounter the following sequence. When a point is added to the alpha complex, a new component is created: the zeroth Betti number $\beta_0$ is increased by 1.
In this study, the alpha complex starts out with all haloes added at $\alpha=0$, and each component corresponds to an individual halo. When adding edges between pairs of points, the components become (super)cluster of haloes. The process of adding edges between points can have two outcomes. Either both points belong to the same, or to different components of the current complex.
When they belong to the same component, the edge creates a new tunnel, increasing the one-dimensional Betti number by 1. The creation of a filament corresponds to the creation of a tunnel. When the edge connects two different components, two (super)clusters merge and $\beta_0$ decreases by 1. When a triangle gets added, it may complete the enclosure of a void or it may close a tunnel. In the first case, a new void is created and $\beta_2$ increases by 1. In the latter case, $beta_1$ is decreased by 1 as the number of tunnels and corresponding filaments decreases. Finally, when a tetrahedron is added, a void is filled. This means that the two-dimensional Betti number $\beta_2$ is lowered by 1.

By assessing the value of $\alpha$ at which a feature is created, the \emph{birth value} $\alpha_b$, and the value at which a feature is destroyed, the \emph{death value} $\alpha_d$, we follow the existence of (super)cluster complexes, of filamentary features and of enclosed voids~\citep[see, e.g.,][]{wilding2021}. The difference between the creation and the destruction value,
\begin{equation}\label{eq:app:persistence}
    \pi \coloneqq \alpha_d-\alpha_b
\end{equation}
denotes the \emph{persistence} $\pi$ of a topological feature. By plotting the values of $\alpha_d$ vs. $\alpha_b$ for all topological features present in a spatial point distribution, for each dimension $d=0,1,\ldots,D$, a highly detailed and idiosyncratic depiction of the topological structure is obtained. These \emph{persistence diagrams} not only quantify the topology in terms of a few summarising parameters, but in terms of diagrams that contain information on every individual topological feature present in the sample point distribution. 

Analogously, we can compute the Betti numbers ($\beta_0$, $\beta_1$ and $\beta_2$) of the distance field as a function of the length scale $\alpha$. The resulting Betti curves (i.e., the Betti numbers for all scales) provide an overview on the complete structure of critical points, and on how the global topology differs for varying scales. While Betti curves inform us of the overall topology (since they measure the total number number of topological features as a function of $\alpha$), persistence diagrams allow us to precisely track at what length scales each of these topological features is formed and for how long they \emph{persist}.

\subsection{Alpha shapes illustrated}
\label{app:alphashapesillustrated}
The full potential and richness of the topological information contained in alpha complexes may be appreciated from the three-dimensional situation. Figs.~\ref{fig:B1_3d_alphashape} and \ref{fig:B2_3d_alphashape_zoom} provide illustrations of the alpha shapes in a 30 $h^{-1}$Mpc thick slice through the P-Millennium simulation, for three different values of $\alpha$.

Fig.~\ref{fig:B1_3d_alphashape} shows the alpha shapes in slices through the $P$-Millennium simulations. They are slices with the full width of $542.16 \ h^{-1}$Mpc of the simulation box, at three
different values of the average halo-halo distance (one third, one half, and two thirds). Fig.~\ref{fig:B2_3d_alphashape_zoom} provides a supplementary view of the simplicial structure, including an additional slice at $\alpha=0.6$ (a quarter of average halo-halo distance). Along with the full slices, the right-hand column shows zoom-ins of 100 $h^{-1}$Mpc width. The zoom-in region is indicated in the full slices in the left-hand column.

Alpha shapes trace the intricate multi-scale structure, diverse morphological features and their mutual connections of points in a web-like network. From the figures we may appreciate the natural means by which alpha shapes follow the multi-scale nature and topological structure of the cosmic web (for reference also see figure~4 in \citet{weygaert2011alpha}). The figures show the emergence of the cosmic web as traced by dark matter haloes. For small values of $\alpha$ (see the top row of Fig.~\ref{fig:B2_3d_alphashape_zoom}) the structure consists of largely disconnected clusters of haloes and their filamentary extensions. Whereas the overall arrangement of clusters amidst filamentary and wall-like extensions is clear, these ``supercluster'' complexes still appear to be mostly singular, isolated features. They are separated by large underdense regions, in which tenuous largely elongated features are visible. These are the denser parts of large-scale filamentary bridges and tendrils. Only at higher values of $\alpha$ we see that the high-density supercluster islands start to get connected via filamentary bridges, connections that become stronger, more elaborate and more structured as more and more features get connected in an emerging pervasive network. It is in particular the central panel of Fig.~\ref{fig:B1_3d_alphashape} that offers the most detailed view of the intricacies of the fully connected spatial network. At even higher values of $\alpha$ the lower density areas start to fill in the cavities in the spatial network. This yields the reverse impression of the structure at lower $\alpha$ values: it is the underdense void regions that turn into ever more isolated and encapsulated cavities. The third panel of Fig.~\ref{fig:B1_3d_alphashape} depicts this phase of the alpha complex.

The zoom-in panels of Fig.~\ref{fig:B2_3d_alphashape_zoom} are centered around a low-density void region. It provides a telling illustration of the substructure of voids and of the signature of the corresponding multi-scale topology in the alpha complex. An almost empty region at the lowest $\alpha$ value, gets gradually structured towards higher $\alpha$ values. At first this concerns isolated clumps, which towards the higher $\alpha$ values get connected in elongated features (third panel). Ultimately, towards the highest $\alpha$ values, these features enclose small subvoids of the initially large void (fourth panel). In short, the four zoom-in panels of Fig.~\ref{fig:B2_3d_alphashape_zoom} provide visually compelling evidence for the ability of alpha shapes to probe the detailed and intricate multi-scale structure of the cosmic web and its components.

Translated into the corresponding Betti number signature, the first panel of Fig.~\ref{fig:B1_3d_alphashape} is marked by the dominance of $\beta_0$, while the central panel corresponds to high values of $beta_1$ and a diminished $\beta_0$ signature due to the merging of supercluster islands into the cosmic web. The third panel has a dominant $\beta_2$ signature, as a result of the increasing abundance of fully enclosed void cavities, while the emergence of their wall-like boundaries implies a substantial decrease of $\beta_1$ due to the disappearance of filament-related tunnels in the corresponding alpha shapes.

\section[Distance field analysis -- Alpha shapes vs. graph methods]{\\ Distance field analysis -- \\ Alpha shapes vs. graph methods}
\label{app:distancefield}
The sampling of the distance field is a well-known and suggestive method for quantifying the cosmic matter distribution probed by the discrete distribution of haloes or galaxies. It is familiar in the context of both halo and clump classification, as well as in a range of techniques for assessing aspects of the spatial distribution of galaxies. The \textsc{Friends-of-Friends} algorithm~\citep{Davis1985, White2010}, and elaborations thereof~\citep{Behroozi2013}, is a widely applied tool for the analysis of $N$-body simulations and identification of haloes.

One of the earliest examples of its exploitation in the context of the large-scale structure of the Universe is that of the percolation analysis by \citet{zeldovich1982}, seeking to establish aspects such as at which scale connected structures form that span the entire volume, or at which scale all points in a sample get connected. A more intricately and uniquely defined structure is that of the Minimum Spanning Tree (MST), a construct stemming from graph theory. It links \emph{all} points in a sample by means of a \emph{graph} with a minimal total length. It has now a range of applications in the analysis of the cosmic galaxy distribution. Bhavsar and collaborators~\citep{Barrow1985,graham1995} introduced MSTs to quantify the filamentary nature of the galaxy distribution. Based on MSTs, \citet{Colberg2007} developed a formalism to identify and classify the web-like distribution of haloes and galaxies, an approach that got further elaborated on by \citet{alpaslan2014b}. Their MST based formalism has been applied for the identification of filaments and void regions in the GAMA survey~\citep{alpaslan2014a}.

Another approach to the use of MSTs concerns an assessment of a range of statistical measures of the MST edge length distribution. \citet{naidoo2020} used this in an attempt to quantify higher order clustering properties. \citet{bonnaire2021} also used aspects of graph theory, and a keenly defined statistical procedure to regularize the MST, to obtain a technique that identifies in particular the dominant filaments in a 2D and 3D galaxy distribution.

In the more general setting of \emph{graph theory}, a range of studies have been using the statistics of edges connecting the points in the spatial particle, halo, or galaxy distribution to extract clustering measures. Statistical moments of the edge length distribution are proposed to quantify aspects of the higher order spatial clustering of the sample points. Expecting that these measures pertain to aspects of their web-like distribution, \emph{network analysis} is obtaining increasing attention~\citep{hongdey2015,hong2016,deregt2018,tsizh2020} in large-scale structure studies.

These applications underline the substantial amount of information contained in the distance field, sampled by the mutual distances between the sample points. Nonetheless, it is through the unique, transparent, geometrically and topologically ordered means by which \emph{alpha shapes} are processing the vast information content of the sample point distribution, and the connections between the points, that the measures extracted can be related to visually accessible and topologically and geometrically meaningful information. They do so in an entirely natural fashion, without the need for ad hoc parameters or fitting to heuristic mathematical structures.

% [Add here]
% If you want to present additional material which would interrupt the flow of the main paper,
% it can be placed in an Appendix which appears after the list of references.

%%%%%%%%%%%%%%%%%%%%%%%%%%%%%%%%%%%%%%%%%%%%%%%%%%

% Don't change these lines
\bsp	% typesetting comment
\label{lastpage}
\end{document}